\documentclass[letterpaper,aps, prl, reprint, superscriptaddress]{revtex4-1}
\pdfoutput=1
\usepackage[T1]{fontenc}
\usepackage{color}
\usepackage{xcolor}
\usepackage{pdfcolmk}
\usepackage{verbatim}
\usepackage{mathtools}
\usepackage{amsmath}
\usepackage{amssymb}
\usepackage{float}
\usepackage{array}
\usepackage{mathrsfs}

\makeatletter

\providecolor{lyxadded}{rgb}{0,0,1}
\providecolor{lyxdeleted}{rgb}{1,0,0}
\DeclareRobustCommand{\lyxsout}[1]{\ifx\\#1\else\sout{#1}\fi}

\usepackage[colorlinks,citecolor=blue,linkcolor=red]{hyperref}

\begin{document}
\title{Realization and detection of Kitaev quantum spin liquid with Rydberg atoms}

 \author{Yi-Hong Chen}
 \thanks{The authors make equal contributions.}
 \affiliation{International Center for Quantum Materials, School of Physics, Peking University, Beijing 100871, China}
 \affiliation{Hefei National Laboratory, Hefei 230088, China}

 \author{Bao-Zong Wang}
 \thanks{The authors make equal contributions.}
 \affiliation{International Center for Quantum Materials, School of Physics, Peking University, Beijing 100871, China}
 \affiliation{Hefei National Laboratory, Hefei 230088, China}

 \author{Ting-Fung Jeffrey Poon}
 \affiliation{International Center for Quantum Materials, School of Physics, Peking University, Beijing 100871, China}
 \affiliation{Hefei National Laboratory, Hefei 230088, China}

 \author{Xin-Chi Zhou}
 \affiliation{International Center for Quantum Materials, School of Physics, Peking University, Beijing 100871, China}
 \affiliation{Hefei National Laboratory, Hefei 230088, China}

\author{Zheng-Xin Liu}
 \affiliation{Department of Physics, Renmin University of China, Beijing 100872, China}

\author{Xiong-Jun Liu}
\thanks{Corresponding author: xiongjunliu@pku.edu.cn}
\affiliation{International Center for Quantum Materials, School of Physics, Peking University, Beijing 100871, China}
\affiliation{Hefei National Laboratory, Hefei 230088, China}
% \affiliation{Collaborative Innovation Center of Quantum Matter, Beijing 100871, China}
\affiliation{International Quantum Academy, Shenzhen 518048, China}
\affiliation{CAS Center for Excellence in Topological Quantum Computation, University of Chinese Academy of Sciences, Beijing 100190, China}

\begin{abstract}
The Kitaev chiral spin liquid has captured widespread interest in recent decades because of its intrinsic non-Abelian excitations, yet the experimental realization is challenging. Here we propose to realize and detect Kitaev chiral spin liquid in a deformed honeycomb array of Rydberg atoms. Through a novel laser-assisted dipole-dipole interaction mechanism to generate both effective hopping and pairing terms for hard-core bosons, together with van der Waals interactions, we achieve the pure Kitaev %spin liquid
model with high precision. The gapped non-Abelian spin liquid phase is then obtained by introducing Zeeman fields. Moreover, we propose innovative strategies to probe the chiral Majorana edge modes by light Bragg scattering and by imagining their chiral motion. %, in which only spin degree of freedom needs to be measured.
Our work %not only facilitates the study of topological matter and non-Abelian anyons, but also
broadens the range of exotic quantum many-body phases that can be realized and detected in atomic systems, and makes an important step toward manipulating non-Abelian anyons.

\end{abstract}

\maketitle

\textcolor{blue}{\em Introduction.}--
Topological order is a captivating concept in condensed matter physics that goes beyond Landau paradigm and can host exotic quasiparticle excitations as anyons \cite{wen2017colloquium}. The Kitaev model on the honeycomb lattice with exactly solvable ground state is an idea platform to prepare for gapped quantum spin liquid (QSL) phases with the Ising-type non-Abelian topological order
%spin liquid is a prominent example of non-Abelian topological order that is exactly solvable, with vortex-bound Majorana zero mode as non-Abelian Ising anyons
\cite{kitaev2006anyons}.
The braiding of the Ising-type anyonic excitations results in
non-Abelian Berry phase %rotations
within the space of topologically degenerate states, which remains robust against local perturbations \cite{nayak2008non,wilczek1990fractional,stern2010non,moore1991nonabelions,wen1991non}.
Furthermore, the promotion of braiding operations to unitary gates is the key of the promising fault-tolerant topological quantum computation \cite{kitaev2003fault,nayak2008non,pachos2012introduction,stern2013topological}.
%Despite the s
Significant efforts %dedicated to
have been made in %exploring Kitaev spin liquids in
solid-state systems \cite{hermanns2018physics,knolle2019field,takagi2019concept,trebst2022kitaev,banerjee2017neutron,do2017majorana,kasahara2018majorana,yokoi2021half,bruin2022robustness}, %conclusive
but the confirmation of %Kitaev spin liquid is obstructed by their intricacy and under debate
%this QSL phase
the Kitaev QSL phases
is still on-going
\cite{baek2017evidence,hentrich2018unusual,bachus2020thermodynamic,leahy2017anomalous,janvsa2018observation,winter2018probing,chern2021sign}.

Alternatively, quantum simulation \cite{georgescu2014quantum,daley2022practical,duan2003controlling} may provide a route to achieve a clean and highly controllable realization of the Kitaev spin liquid. The dynamical simulation approaches were proposed recently related to the idea of Floquet spin liquid \cite{bookatz2014hamiltonian,po2017radical}, including a digital simulation through Rydberg atoms \cite{lukin2022non} and a dynamical simulation \cite{Monika020329} on the optical lattice.
The Rydberg atom platform is particularly suitable for simulating the spin models \cite{browaeys2020many,glaetzle2014quantum,nguyen2018towards,celi2020emerging,mazza2020vibrational,lee2023landau,nishad2023quantum,kunimi2023proposal}, with Rydberg atoms in optical tweezer arrays  \cite{barredo2016atom,endres2016atom,kim2016situ,barredo2018synthetic,de2019defect} resembling lattice spins, and the dipole-dipole and van der Waals (vdW) interactions inherently produce XY and Ising interactions \cite{browaeys2020many,saffman2010quantum}. This capability of simulating quantum spin systems has been verified in numerous experiments, including quantum Ising model \cite{labuhn2016tunable,bernien2017probing,keesling2019quantum,scholl2021quantum,ebadi2021quantum,bluvstein2021controlling}, XY model \cite{barredo2015coherent,chen2023continuous}, one-dimensional bosonic symmetry-protected topological phase \cite{de2019observation}, and abelian quantum spin liquid \cite{semeghini2021probing}. Nevertheless, the simulation of generic spin interactions could be challenging.
To this end, the laser-assisted dipole-dipole interaction (LADDI) technique in Rydberg atom array is introduced in our recent works \cite{PhysRevA.106.L021101,poon2023fractional,zhou2022exact}, extending Raman-engineered single-particle couplings studied in optical lattices \cite{aidelsburger2013realization,miyake2013realizing,liu2014realization,aidelsburger2015measuring,wu2016realization,liu2016chiral,song2018observation,lu2020ideal,wang2021realization}. This technique provides a flexible tool to precisely control spin-exchange interactions through lights, equipping %. Consequently, it equips
us with the capability to realize more intricate and exotic quantum spin models.

In this letter, we propose a novel scheme to precisely realize and probe the Kitaev spin liquid in a deformed honeycomb array of Rydberg atoms.
The %Kitaev Ising
Kitaev exchange interactions are decomposed into the hopping and pairing terms of the hard-core bosons, which are realized based on a fundamentally new type of LADDI proposed here. Together with vdW interaction, we show the Kitaev spin liquid model can be realized with a high tunability and precision. We further propose the feasible schemes to detect the non-Abelian spin liquid, including the bulk gap and chiral edge state, through dynamical response and unidirectional transport measurements. Our work opens a promising avenue for exploring the quantum spin liquids and their potential applications.

\begin{figure}[htbp]
\includegraphics[width=1.0\columnwidth]{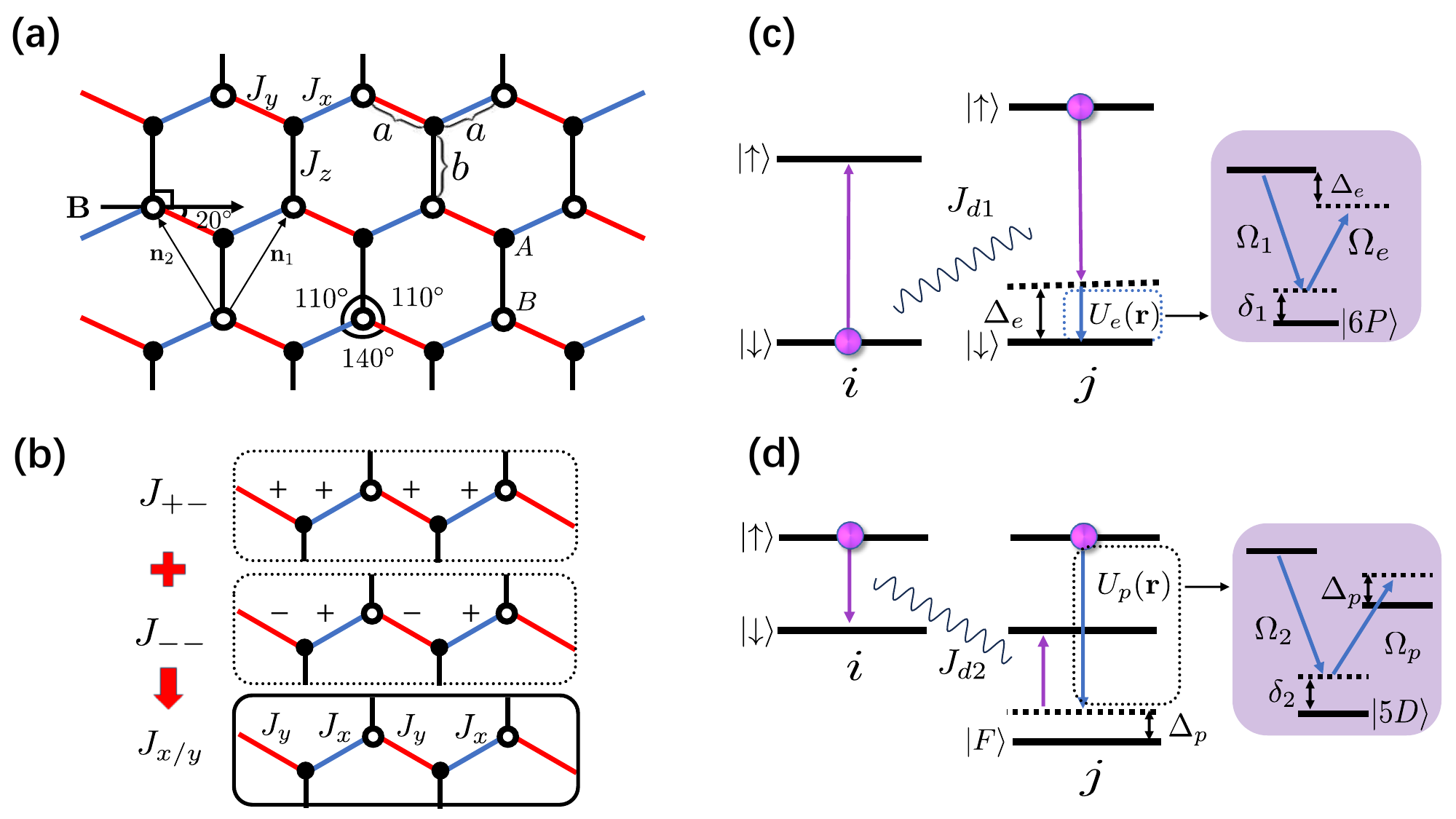}
\caption{\label{Fig1}Sketch of the model and realization. (a) The transformed lattice configuration of the Kitaev honeycomb model, which is realized by Rydberg atoms trapped in the optical tweezers.
The angle between bonds and bond lengths deviates slightly from the ideal honeycomb lattice.
% An in-plane magnetic field is introduced to define the quantization axis of angular momentum.
(b) The alternating $J_x$, $J_y$ configuration consists of a uniform hopping term $J_{+-}$ and a staggered pairing term $J_{--}$ with the same absolute value. (c-d) Illustration of LADDIs. Inset: the two-photon Raman processes (details in supplementary material). (c) The hopping process $J_{+-}$, where Raman potential $U_e=\Omega^{\star}_{1}\Omega_{e}/\delta_1$ compensates the energy offset, with $\delta_1$ representing the detuning of the two Raman lasers to the intermediate state. (d) The pairing interaction $J_{--}$, driven by the Raman potential $U_p=\Omega^{\star}_2\Omega_p/\delta_2$, causes the same spin flip process on both sites.}
% occurs when one Rydberg atom is coupled to the intermediate state $|F\rangle$ by the Raman potential $U_p$, followed by the bare dipole-dipole interaction $J_{d2}$ couples to the flipped spin states. }	
\end{figure}

\textcolor{blue}{\em Model.}--
The paradigmatic Kitaev model is an ideal platform %spin liquid is a paradigmatic model
%which exhibits
to prepare for non-Abelian anyonic excitations. The model under consideration is defined by a spin-1/2 system on a honeycomb array, as depicted in Fig. \ref{Fig1}(a), and described by the Hamiltonian given by:

\begin{equation}\label{model1}
	H=-\sum_{\langle ij \rangle_\gamma}J_{\gamma}s^{\gamma}_{i}s^{\gamma}_j-\sum_{j,\gamma}h_{\gamma}s_j^{\gamma},\,\gamma=x,y,z.
\end{equation}
Here $\langle ij \rangle_\gamma$ labels $\gamma=(x,y,z)$-type, %$y$-type and $z$-type bond,
the summation is taken over all the bonds on the honeycomb lattice, and $J_\gamma$ denotes the Ising coupling strength for spin components $s^{\gamma}_{i(j)}$ of the $\gamma$-bond connecting sites $\mathbf{r}_i$ and $\mathbf{r}_j$ [Fig. \ref{Fig1}(a)]. %The spin-$1/2$ operators $s^{\gamma}_{i}$ are located at the lattice site $\mathbf{r}_i$, and the Hamiltonian term is defined over the two ends of the $\gamma$ bond connecting sites $\mathbf{r}_i$ and $\mathbf{r}_j$, as shown in Fig. \ref{Fig1}(a).
This model for $h_{\gamma}=0$ is exactly solvable by reduction to free majorana fermions in static $\mathbb{Z}_2$ gauge field, featuring three gapped A phases and one gapless B phase~\cite{kitaev2006anyons}. %(see Fig. \ref{Fig3}). %Applying Zeeman term
Applying the Zeeman field breaks time reversal symmetry and %can
drives the B phase into a non-Abelian spin liquid. %with vortex majorana zero modes.

Nonetheless, directly simulating these bond-dependent Ising interactions with Rydberg atom array is out of reach, since the intrinsic dipole-dipole and vdW interactions between Rydberg atoms are of XXZ type~\cite{browaeys2020many,saffman2010quantum}. %XY and Z-Ising types.
Our key idea is to %We overcome such difficulty by first
decompose these Ising interactions into hopping ($J_{+-}$) and pairing ($J_{--}$) terms, and then realize them separately, in the hard-core boson language \footnote{$J_{+-}^{\gamma}$ ($J_{--}^{\gamma}$) denote hopping (pairing) term across $\gamma$ bond.}
\begin{equation}\label{Hrow}
\left\{
\begin{aligned} \sum_{\langle ij \rangle_{x}}J_xs^x_{i}s^x_j = \sum_{\langle ij \rangle_{x}} \left(J^x_{+-}b_i^{\dagger} b_j+J^x_{--}b_i b_j\right )  +h.c., \\
\sum_{\langle ij \rangle_y}J_y s^y_i s^y_j = \sum_{\langle ij \rangle_y } \left(J^y_{+-}b_i^{\dagger} b_j+J^y_{--}b_i b_j\right ) +h.c.,
\end{aligned}
\right.
\end{equation}
Here the hard-core bosons are defined as $b_j=s_j^-$ with $b_j^2=0$, the hopping and pairing terms along ($x,y$)-bonds are the same (opposite) for $x$ ($y$) bonds, namely, $J^{x}_{+-}=J^{x}_{--}=J_x/4$ and $J^{y}_{+-}=-J^{y}_{--}=J_y/4$ [see Fig.~\ref{Fig1}(b)]. %Thus the %$J_x$($J_y$) Ising couplings require that
%namely, the hopping and pairing terms are the same (opposite) for $x$ ($y$) bonds [see Fig.~\ref{Fig1}(b)]. %, while $J_{--}$ exhibits a staggered pattern along the row [Fig.~\ref{Fig1}(b)].
Unlike the hopping term arising from dipole-dipole interaction, the pairing term is intrinsically not present for Rydberg atoms. We present here an innovative scheme to solve the challenge and realize both the hopping and pairing terms through a novel type of LADDI. Together with a well engineered vdW interaction to realize the $z$-bond interaction, we achieve the model~\eqref{model1} with high precision.

\textcolor{blue}{\em Experimental scheme.}--We demonstrate the realization of controllable hopping and pairing interactions in a 2D Rydberg atom array, composed of $^{87}$Rb atoms trapped in an optical tweezer array with the deformed honeycomb lattice configuration. For our purpose, the bond angles of the lattice are deformed to $110^{\circ}$, %instead of $120^{\circ}$,
and the $x,y$ ($z$) bond lengths are $a$ ($b$), as shown in Fig.\ref{Fig1}(a).
A real in-plane magnetic field is introduced to define the "quantization axis" perpendicular to the $z$-bond and at an angle $\theta=20^{\circ}$ to the $x,y$-bond. The spin states are represented by $\left|\downarrow\right\rangle =|n_D D_{\frac{5}{2}},m_{J}=\frac{5}{2}\rangle$ and $\left|\uparrow\right\rangle =|n_P P_{\frac{3}{2}},m_{J}=\frac{3}{2}\rangle$.

The hopping term $J_{+-}$, which %couples
transforms $\left|\downarrow \right\rangle_i\left|\uparrow\right\rangle_j$ %and
into $\left|\uparrow\right\rangle_i \left|\downarrow \right\rangle_j$, can be constructed through a laser-assisted process, here $i$ and $j$ label two nearest neighbor sites on the $x$- or $y$-bonds.
As shown in Fig.~\ref{Fig1}(c), the bare dipole-dipole interaction $J_{d1}$ which couples $\left|\downarrow\right\rangle_i \left|\uparrow\right\rangle_j\rightarrow\left|\uparrow\right\rangle_i \left|\downarrow\right\rangle_j$ is suppressed by a detuning $\Delta_e=(E_{\uparrow}-E_{\downarrow})_j-(E_{\uparrow}-E_{\downarrow})_i$.
The detuning is induced by introducing a site-dependent energy shift to the spin-up state $E_{\uparrow,i}=E_P+\Delta_i$. In the presence of Raman potential $U_{e}(\mathbf{r})e^{i\Delta_e t}$, this energy offset is compensated, enabling the laser-assisted exchange process. Therefore, by applying energy offset $\Delta_e$, the exchange couplings are completely controllable by the lasers.
From a standard perturbation theory \footnote{See more details in Supplementary material about the laser assisted dipole-dipole interaction, numerical estimate of parameters, and the detection schemes.}\label{supplement}, we obtain
\begin{equation}\label{exchange eq}
J_{+-}= \frac{U_e(\mathbf{r}_{j})}{\Delta_e}J_{d1}+O((J_{d1},U_e)^4),
\end{equation}

The pairing term $J_{--}$ transforms $\left|\uparrow\right\rangle_i\left|\uparrow\right\rangle_j$ % and
to $\left|\downarrow\right\rangle_i \left|\downarrow\right\rangle_j$, and is constructed through another laser-assisted process through an intermediate state $|F\rangle= |n_F F_{7/2},m_{J}=7/2\rangle$, as illustrated in Fig. \ref{Fig1}(d). The dipole-dipole interaction $J_{d2}$, which couples $\left|\uparrow\right\rangle_i\left|F\right\rangle_j$ to $\left|\downarrow\right\rangle_i\left|\downarrow\right\rangle_j$, takes effect when the detuning $\Delta_{p}=(E_{\downarrow}-E_{\uparrow})_i+(E_{\downarrow}-E_{F})_j$ is compensated by the Raman potential $U_{p}(\mathbf{r})e^{i(E_{\uparrow}-E_F-\Delta_p)t}$, giving rise to the process $\left|\uparrow\right\rangle_i\left|\uparrow\right\rangle_j\rightarrow\left|\uparrow\right\rangle_i\left|F\right\rangle_j\rightarrow\left|\downarrow\right\rangle_i\left|\downarrow\right\rangle_j$. We obtain from perturbation %\cite{suzuki1983degenerate}, we obtain
\begin{equation}\label{pairing eq}
    J_{--}=-\frac{U_p(\mathbf{r}_j)}{\Delta_p}J_{d2}+O((J_{d2},U_p)^4),
\end{equation}
under the resonant Raman coupling regime~\footnotemark[2].

\begin{figure}[htbp]
\includegraphics[width=1.0\columnwidth]{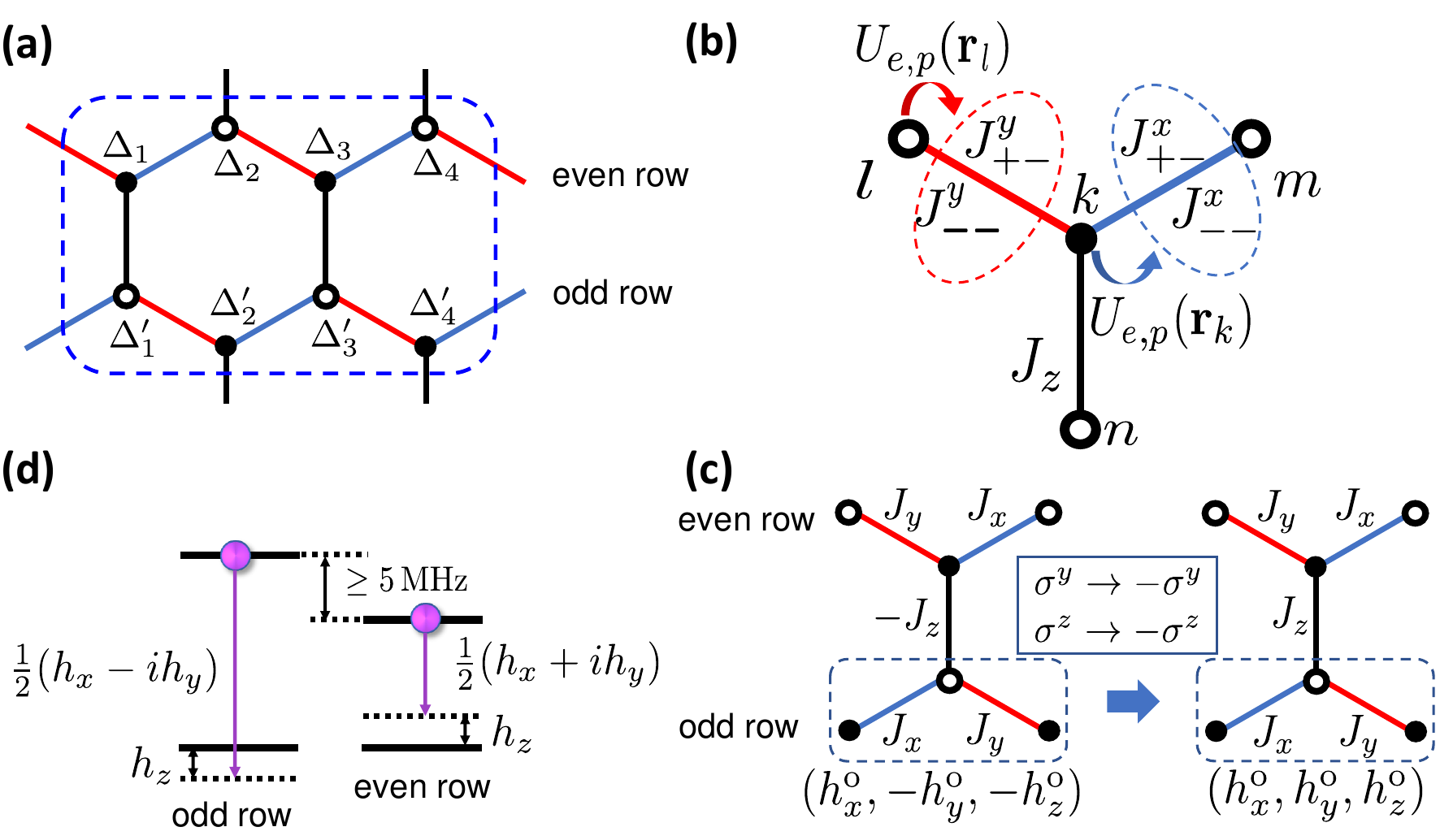}
\caption{\label{Fig2}Detailed configuration of the experimental proposal. (a) The on-site energy shifts, The detunings $\Delta_i$ and $\Delta'_i$ at even and odd rows are set to be different. (b) The coupling at a bond is determined only by the Raman potential on one side.
% For example, the coupling $J_{+-(--)}$ at the $\langle kl\rangle$ bond is assisted by
Raman potential $U_{e,p}(\mathbf{r}_{l})$ ($U_{e,p}(\mathbf{r}_{k})$) have frequencies corresponding to $\Delta_{e,p}$ on $\langle lk\rangle$ ($\langle km\rangle$) bond.
(c) A unitary transformation that transforms $\sigma_{y,z}$ in odd rows to $-\sigma_{y,z}$, resulting in $J_z$ changing sign but $J_{x,y}$ remaining unchanged. It also changes the sign of $h_{y,z}$ in odd rows (labelled $h_{y,z}^{\mathrm{o}}$). (d) The Zeeman terms $h_{y,z}$ are opposite in even and odd rows, which are induced independently by microwaves. %with different frequencies seperated by at least $5\,\text{MHz}$.
}
\end{figure}

The sign configurations of $J^{x,y}_{+-}$ and $J^{x,y}_{--}$ along $(x,y)$-bonds
can be achieved via on-site tuning of the Raman potentials. With the site-dependent energy shift $\Delta_i \ (\Delta_i')$, which has a periodicity of every eight sites [Fig.~\ref{Fig2}(a)], achievable through local tuning of optical tweezers, we can apply different Raman potential $U_{e,p}$ to different bonds separately by matching the Raman potential frequency with the corresponding detuning $\Delta_{e,p}$ [Fig. \ref{Fig2}(b)].
The phases of the Raman potential $U_e$ are chosen to ensure a uniform sign for $U_e/\Delta_e$, resulting in %that
$J^x_{+-} = J^y_{+-}$.  Meanwhile, the signs of $U_p$ exhibit a distinctive staggered pattern along the row, giving rise to the  relation of $J^x_{--} = -J^y_{--}$. This yields the $J_{x,y}$ exchange terms for the Kitaev model, %spin liquid,
as depicted in Fig.~\ref{Fig1}(c).

The $J_{z}$ term arises from the intrinsic vdW interaction of Rydberg atoms, characterized by $V_{\sigma\sigma'}\propto n_{\sigma}n_{\sigma'}$, where $n_{\sigma}$ denotes the occupation number at state $|\sigma = \uparrow,\downarrow\rangle$. From %the spin operator
$s^z=(n_{\uparrow}-n_{\downarrow})/2$, we obtain the $J_z$ term by
% -\sum_{\langle ij \rangle_{z}}J_z s_i^z s_j^z
\begin{align} H_{\mathrm{inter-row}} = &-\sum_{\langle ij \rangle_{z}}J_z(n_{i,\uparrow}-n_{i,\downarrow})(n_{j,\uparrow}-n_{j,\downarrow})/4.
\end{align}
% where $V_{\alpha\beta}$ denotes the vdW interaction between the $\alpha$ state on site $\mathbf{r}_i$ and the $\beta$ state on site $\mathbf{r}_j$.
Here $J_z=V_{\downarrow\uparrow}+V_{\uparrow\downarrow}-V_{\downarrow\downarrow}-V_{\uparrow\uparrow}$. Note that the lattice sites on even and odd rows have different energy offsets $\Delta_i$ and $\Delta^{\prime}_i$, respectively [Fig.~\ref{Fig2}(a)]. Thus the interactions $J_{+-}$ and $J_{--}$ at $z$ bonds are off-resonant and negligible. Only the $J_{z}$-coupling remains along the $z$ bonds.

\begin{figure}[htbp]
\begin{centering}
\includegraphics[width=\columnwidth]{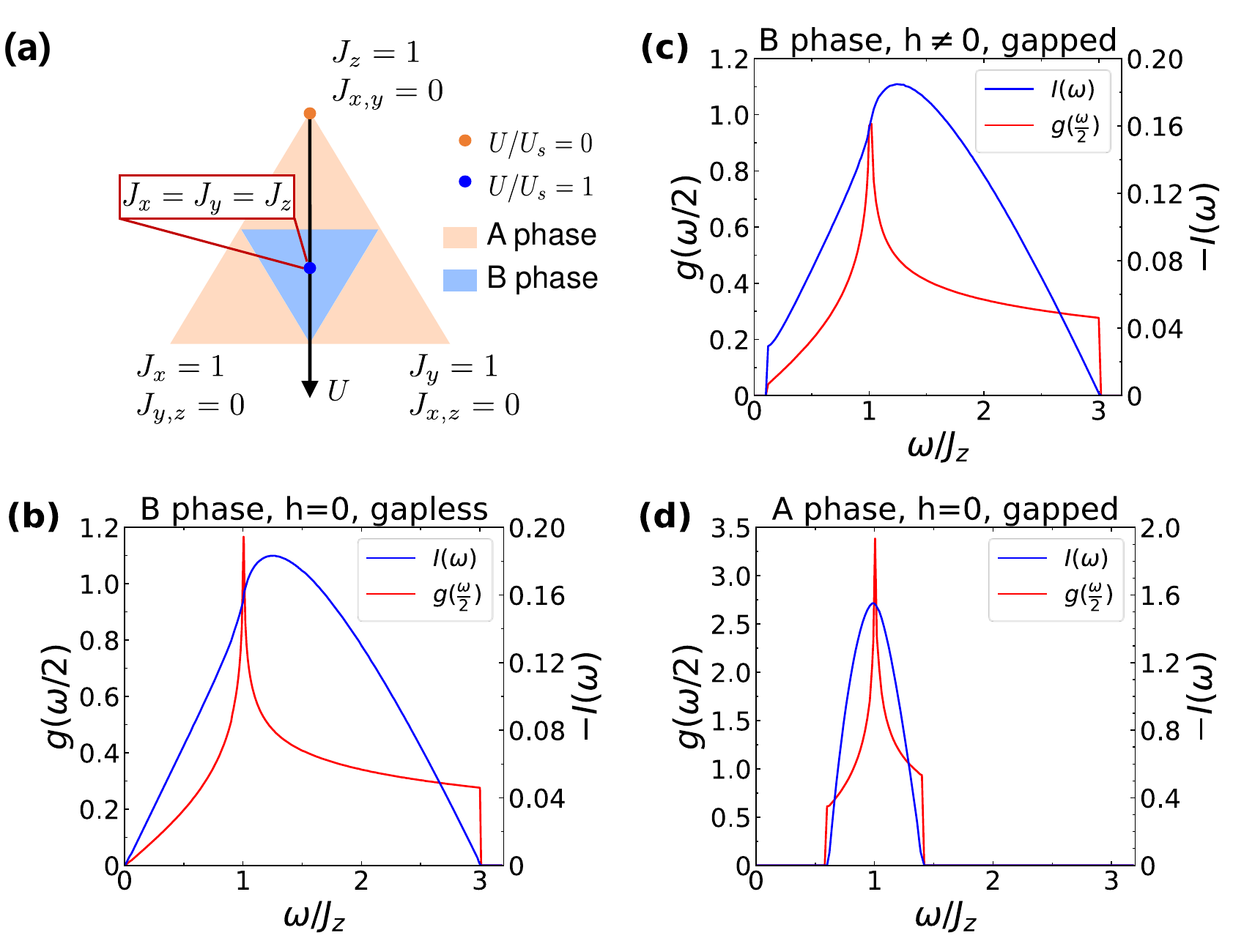}
\par\end{centering}
\centering{}\caption{Phase diagram and detection of the bulk spectra. (a) Ground state phase diagram of the Kitaev honeycomb model.
The parameters evolve along the black line by tuning the strength of Raman potential $U$.
% By tuning the strength of Raman potential $U$ from $0$ to $U_s$ (corresponding to the symmetry point $J_x=J_y=J_z$), the system evolves from a gapped abelian phase (orange point) to the gapless non-Abelian phase (blue point).
(b)-(d)
% Detecting the majorana gap of Kitaev spin liquid by applying two-spin perturbation and measuring the response.
Two-spin response $I(\omega)$ and underlying density of states $g(\omega/2)$.
% whose common lower cutoff gives the bulk gap $2\Delta$.
(b)-(c) The Kitaev B phase \cite{kitaev2006anyons} with the parameters $J_x=J_y=J_z$. The gapless B phase acquires a topological gap when Zeeman term is applied.
% which is gapless without Zeeman term, and acquires a topological gap when Zeeman term is applied.
(d) The abelian Kitaev A phase at $J_x=J_y=J_z/5$.
% which is gapped without Zeeman term.
% in (b) and (c), and  in (d).
}\label{Fig3}
\end{figure}

Our proposal offers high tunability for simulating the Kitaev spin liquid, encompassing both Kitaev A and B phases~\cite{kitaev2006anyons}. By adjusting the amplitude of the Raman potential $U_{e,p}$ while keeping $J_z$ constant, the parameters $J_{x,y}$ can be continuously tuned from $J_x=J_y=0$ in the A phase to $J_x=J_y=J_z$ in the B phase, as illustrated by the black line in Fig. \ref{Fig3}(a).

\textcolor{blue}{\em Zeeman terms.}--The non-abelian anyons emerge in phase B by applying Zeeman terms $h_{x,y,z}$ to open a topological gap. The naturally realized $J_z$ is typically ferromagnetic ($J_z > 0$). %which makes it vulnerable to perturbations.
%To enhance the robustness of the system against such perturbations, To enhance system robustness, we aim to
We can physically transform it into antiferromagnetic interactions by properly applying Zeeman terms.
First, we map $\sigma_{y,z}$  to $-\sigma_{y,z}$ in odd rows through a unitary transformation [Fig. \ref{Fig2}(c)], causing $J_z$ to change sign while leaving $J_{x,y}$ unchanged. This transformation also flips the Zeeman field
$h_{y,z}$ in odd rows to $-h_{y,z}$. Then, in order to maintain the uniformity of the Zeeman field, %after the transformation,
the originally applied Zeeman components $h_{y,z}$ should have opposite signs in even and odd rows, which can be implemented as follows.

The $h_z$-term is directly obtained by setting detuning ($h_z^e$) for the transition induced by Raman potential $U_p$ between $\left|\downarrow\right\rangle_i\left|\downarrow\right\rangle_j$ and $\left|\uparrow\right\rangle_i\left|\uparrow\right\rangle_j$. A staggered configuration of $h_z$ is resulted by tuning opposite detunings in even and odd rows [Fig.~\ref{Fig2}(d)]. The $h_{x,y}$ terms are induced by microwaves that drive spin-flip transitions. Similarly, the $h_y$ terms on odd and even rows can be controlled independently by microwaves with different frequencies, hence a staggered configuration of $h_y$ can be feasibly achieved by tuning the coupling phases [Fig.~\ref{Fig2}(d)]. More details are presented in Supplementary Material~\footnotemark[2]
% $h_y$ is opposite in even and odd rows, and becomes uniform after unitary transformation introduced above.

\textcolor{blue}{\em Numerics.}--We take $n_P=44$, $n_D=42$ and $n_F=40$ to exemplify the high precision of the realization. On $(x,y)$-bonds we have $J_{x,y}\simeq-4${\rm MHz} for $a=4.7\mu$m. By choosing the quantization axis and lattice geometry shown in Fig. \ref{Fig1}(a), we find that on the $z$-bond one can set $J_z\simeq4.0${\rm MHz} for $b=4.2\mu$m, while on the $x,y$ bonds it is quite small $|J_z|<0.16{\rm MHz}$~\footnotemark[2], giving a nearly pure Kitaev model. Further, applying the Zeeman term $h_{x,y,z}\sim0.5{\rm MHz}$ leads to a topological gap in B phase $\sim 0.4\,\mathrm{MHz}$ \cite{zhu2018robust}. Such interactions are sufficiently large for Rydberg atom
lifetime $\tau\gtrsim 60\,\mu\mathrm{s}=240|J_{x,y,z}|^{-1}$.

\textcolor{blue}{\em Detection of the bulk spectrum.}--\label{detectgap}We now turn to the detection of the bulk spectra of Majorana fermions $c_i$ \cite{kitaev2006anyons}.
% that reside on the same honeycomb lattice, derived through a transformation from spin as  proposed by Kitaev
The gap can be detected through a minimal perturbation approach utilizing ${c_i c_j}_{\langle ij\rangle_{\gamma}}$ to excite the Majorana fermions,  which is equivalent to ${s^{\gamma}_{i} s^{\gamma}_{j}}_{\langle ij\rangle_{\gamma}}$ in the spin representation. The two-spin perturbation Hamiltonian given by $\Delta H=\sum_{\langle ij\rangle_{\gamma=x,y}} \Delta J  s^{\gamma}_{i} s^{\gamma}_{j}$ is applied shorter than typical inverse gap ($\sim2.5\,\mu\mathrm{s}$) in B phase with Zeeman field.
% \begin{equation}\label{gap perturbation}
% H_R = \sum_{\langle ij\rangle_{x}} J_R  s^{x}_{i} s^{x}_{j}+\sum_{\langle ij\rangle_{y}} J_R  s^{y}_{i} s^{y}_{j}.
% \end{equation}
Experimentally, this is achieved by varying $J_{x,y}$ through Raman potentials. The response is determined by retarded function $iF(t)=\langle [\Delta H(t),\Delta H(0)]\rangle\theta(t)$, where the average is taken over ground state. The Fourier transformation $I(\omega)\equiv\int_{0}^{\infty}dte^{i\omega t}F(t)$ yields \cite{knolle2014raman}
\begin{equation}\label{gap response}
    I(\omega)=-4\pi \sum_{\mathbf{q}}\delta(\omega-4|s_{\mathbf{q}}|)(\mathrm{Im}[h_{\mathbf{q}}s_{\mathbf{q}}^*]/|s_{\mathbf{q}}|)^2
\end{equation}
for $\omega>0$, and $I(-\omega)=-I(\omega)$. Here $s_{\mathbf{q}}=(J_z+J_x e^{i\mathbf{q}\cdot\mathbf{n}_1}+J_y e^{i\mathbf{q}\cdot\mathbf{n}_2})/4$ and $h_{\mathbf{q}}=\Delta J(e^{i\mathbf{q}\cdot\mathbf{n}_1}+e^{i\mathbf{q}\cdot\mathbf{n}_2})/4$. The dispersion of majorana fermion is $2|s_{\mathbf{q}}|$, so $I(\omega)$ reflects density of states $g(\omega/2)$. The bulk majorana gap $2\Delta$ can be read out from the lower cutoff frequency of $I(\omega)$, as shown in Fig. \ref{Fig3}(b,d). When the Zeeman field %term
is applied, B phase acquires a topological gap. Considering the low energy effective Hamiltonian, we find again $I(\omega)\propto g(\omega/2)$ with a modified factor~\footnotemark[2], detecting the bulk topological gap, as plotted in Fig. \ref{Fig3}(c).

\begin{figure}[tbp]
	\begin{centering}
		\includegraphics[width=\columnwidth]{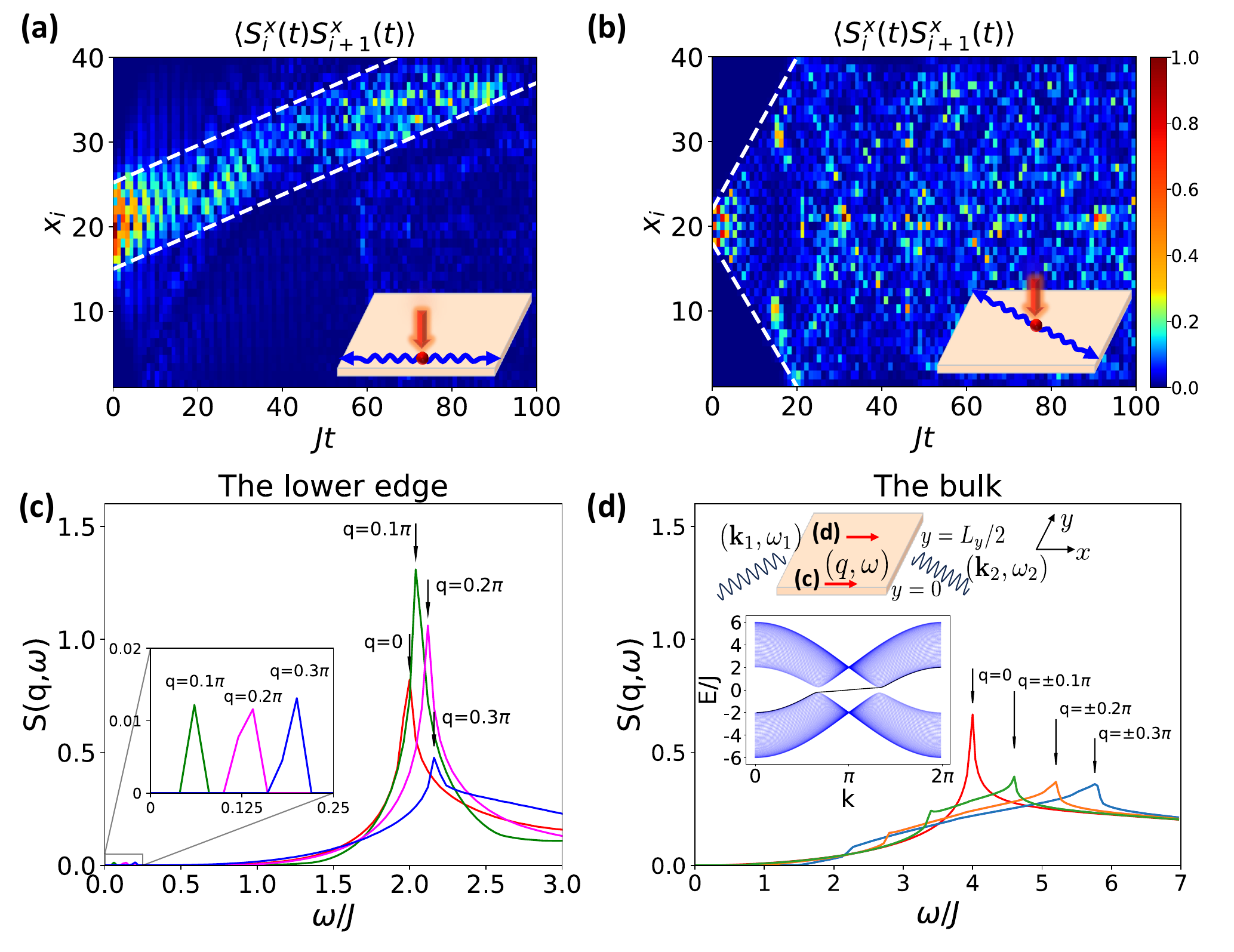}
		\par\end{centering}
\centering{}
\caption{Identifying the chiral edge modes. (a)-(b) Applying a perturbation on an $x$ bond (red point) and then measuring the propagation of spin correlation (blue wavy line) on a $40\times40$-site system. (a) The perturbation is imposed at the edge. Unidirectional propagation is observed. (b) The perturbation is applied in the bulk, and it rapidly diffuses into the entire system symmetrically. (c-d) The dynamical structure factor. (c) The $y=0$ edge region shows peaks at both low and high frequencies at positive transfer momenta.
% The group velocity $v_f=\Delta \omega_{\mathrm{peak}}/\Delta q>0$.
(d) The bulk region only shows peaks in high frequency but symmetric in both positive and negative momenta. Inset: Schematic of light Bragg scattering applied on edge or bulk (red arrow) and energy spectrum with lower edge state (black line)}\label{Fig4}
\end{figure}

\textcolor{blue}{\em Detection of chiral edge modes.}--The gapped Kitaev QSL hosts chiral Majorana edge modes, which emerge in the natural boundaries of the Rydberg atom array, and are a direct measurement of this exotic topological order.
The high tunability of the present Rydberg platform enables unique schemes for the observation. %However, the neutral nature of the Majorana modes hinders direct detection.
We propose below two innovative detection strategies to identify and characterize the Majorana chiral edge modes. %It is worth noting that these methods can be applied to detect the edge modes in other system as well.

The first strategy is to image the chiral motion of the Majorana edge modes. The chiral edge state carries unidirectional %particle
quasiparticle flow, in sharp contrast to the isotropic flow of the bulk states. We show that such sharp difference can be identified by examining merely the spin correlation dynamics. As illustrated in the inset of Fig.~\ref{Fig4}(a,b), we apply a pulse locally to suddenly tune $J_x\rightarrow J_x+\delta J_x$ on a specific $x$ bond. The pulse couples to ${s_i^x s_j^x}_{\langle ij\rangle_x}=-i c_i c_j/4$ and locally excites a two-particle wavepacket. We set the pulse strength as $\delta J_x=0.5$ and the pulse duration as $\delta t=0.1 J_x^{-1}$.
The pulse consists of various frequencies, and Majorana fermions can be excited within a range of energy bands, and then move according to local band dispersion. Subsequently, we measure the motion via spin-correlation on the $x$ bond, denoted as $\langle s_i^x(t) s_j^x(t)\rangle_{\langle ij\rangle_x}$, in two opposite directions indicated by blue arrows in Fig.~\ref{Fig4}.  When the pulse is applied exclusively to the edge [Fig. \ref{Fig4}(a)], the excited spin correlation evolves unidirectionally along the edge, implying an inherent consequence of the chiral nature of edge states, with group velocity given by the edge energy spectrum. In contrast, if the pulse is applied within the bulk, the spin correlation rapidly diffuses symmetrically throughout the entire system in Fig. \ref{Fig4}(b). This sharp contrast provides a clear identification of the chiral Majorana modes in the topological QSL phase.

The second strategy is that we extend the light Bragg scattering technique for detecting Dirac fermion edge modes~\cite{liu2010quantum,stanescu2010topological,goldman2012detecting,buchhold2012effects} to the present chiral Majorana edge modes. Shining two lasers with wave vectors $\mathbf{k}_1(\omega_1)$ and $\mathbf{k}_2(\omega_2)$ on one row of the lattice (see Fig.~\ref{Fig4}(d) inset) induces a Raman process characterized by $\Delta H^{\prime} (\mathbf{q}) = \sum_{\langle ij\rangle_{x}} \Delta J^{\prime} e^{i \mathbf{q}\cdot\mathbf{r}}  s^{x}_{i} s^{x}_{j}$.
Here $\boldsymbol{q}=\boldsymbol{k}_1-\boldsymbol{k}_2 = q \hat{e}_x$ due to momentum conservation along $x$ direction. We obtain the dynamical structure factor $S(q, \omega)$~\footnotemark[2] for the Raman process induced by the shinning lasers. Fig.~\ref{Fig4}(c-d) depicts the numerical simulation. When the perturbation is applied at the $y=0$ edge, two groups of peaks are observed in Fig.~\ref{Fig4}(c): one group of peaks are located at lower frequencies, magnified in the inset, while the other group appears at higher frequencies. The low-frequency peaks exhibit the relationship $\omega=+ q v_f$ when $q>0$, arising from the transition between edge states. In comparison, the high-frequency peaks originate from the transition from bulk states around the smeared van Hove singularity at $k_x=\pi$ to edge states, and show clearly the chirality of the edge modes.
% also exhibiting increasing central frequency $\omega$ with increasing momentum change $q$.
For $q<0$, no resonant peak is observed since such edge states are all occupied~\footnotemark[2].
In stark contrast, the dynamical structure factors for shinning the bulk shows only finite-frequency and symmetric peaks [Fig. \ref{Fig4}(d)], implying that the bulk is gapped and non-chiral. %the absence of transitions between edge modes. Moreover, the curves exhibit symmetry for both positive and negative $q$ values.
With these measurements the chiral Majorana edge mode can be identified.

\textcolor{blue}{\em Conclusion and outlook.}--We have proposed to realize and detect the Kitaev non-Abelian chiral spin liquid in deformed honeycomb array of Rydberg atoms with feasibility. A new LADDI mechanism is introduced to generate hopping and pairing terms in the hard-core boson representation, %picture,
yielding the Kitaev exchange interactions in the $x$- and $y$-bonds. Together with engineering the van der Waals interactions, we realized the pure Kitaev
model with high precision. The gapped non-Abelian spin liquid phase with Ising-type anyons is further realized. %Majorana modes.
% The versatility of laser-assisted interactions allows us to explore both the Kitaev A and B phases and implement local modulation.
Finally, %we proposed
based on measuring spin observables and the dynamical response, we proposed detection schemes
%and unidirectional transport
to identify the gapless and chiral Majorana edge modes. %of the Kitaev spin liquid.
This work opens an avenue in simulating non-Abelian topological orders using Rydberg atoms. Future interesting issues that can be examined with the present realization include further exploring the Majorana related exotic physics like the half-quantized thermal Hall effect, and the non-Abelian statistics of the vortex Majorana modes based on the local tunability unique to the Rydberg atom array, which shall facilitate the applications to practical topological quantum computation.

\textcolor{blue}{\em Acknowledgments.}--This work was supported by National Key Research and Development Program of China (2021YFA1400900 and 2022YFA1405301), the National Natural Science Foundation of China (Grants No. 11825401, No. 12261160368, No. 11974421, and No. 12134020), the Innovation Program for Quantum Science and Technology (Grant No. 2021ZD0302000), and the Strategic Priority Research Program of the Chinese Academy of Science (Grant No. XDB28000000).

\bibliography{ref}

\begin{thebibliography}{80}%
\makeatletter
\providecommand \@ifxundefined [1]{%
 \@ifx{#1\undefined}
}%
\providecommand \@ifnum [1]{%
 \ifnum #1\expandafter \@firstoftwo
 \else \expandafter \@secondoftwo
 \fi
}%
\providecommand \@ifx [1]{%
 \ifx #1\expandafter \@firstoftwo
 \else \expandafter \@secondoftwo
 \fi
}%
\providecommand \natexlab [1]{#1}%
\providecommand \enquote  [1]{``#1''}%
\providecommand \bibnamefont  [1]{#1}%
\providecommand \bibfnamefont [1]{#1}%
\providecommand \citenamefont [1]{#1}%
\providecommand \href@noop [0]{\@secondoftwo}%
\providecommand \href [0]{\begingroup \@sanitize@url \@href}%
\providecommand \@href[1]{\@@startlink{#1}\@@href}%
\providecommand \@@href[1]{\endgroup#1\@@endlink}%
\providecommand \@sanitize@url [0]{\catcode `\\12\catcode `\$12\catcode
  `\&12\catcode `\#12\catcode `\^12\catcode `\_12\catcode `\%12\relax}%
\providecommand \@@startlink[1]{}%
\providecommand \@@endlink[0]{}%
\providecommand \url  [0]{\begingroup\@sanitize@url \@url }%
\providecommand \@url [1]{\endgroup\@href {#1}{\urlprefix }}%
\providecommand \urlprefix  [0]{URL }%
\providecommand \Eprint [0]{\href }%
\providecommand \doibase [0]{https://doi.org/}%
\providecommand \selectlanguage [0]{\@gobble}%
\providecommand \bibinfo  [0]{\@secondoftwo}%
\providecommand \bibfield  [0]{\@secondoftwo}%
\providecommand \translation [1]{[#1]}%
\providecommand \BibitemOpen [0]{}%
\providecommand \bibitemStop [0]{}%
\providecommand \bibitemNoStop [0]{.\EOS\space}%
\providecommand \EOS [0]{\spacefactor3000\relax}%
\providecommand \BibitemShut  [1]{\csname bibitem#1\endcsname}%
\let\auto@bib@innerbib\@empty
%</preamble>
\bibitem [{\citenamefont {Wen}(2017)}]{wen2017colloquium}%
  \BibitemOpen
  \bibfield  {author} {\bibinfo {author} {\bibfnamefont {X.-G.}\ \bibnamefont
  {Wen}},\ }\bibfield  {title} {\bibinfo {title} {Colloquium: Zoo of
  quantum-topological phases of matter},\ }\href@noop {} {\bibfield  {journal}
  {\bibinfo  {journal} {Reviews of Modern Physics}\ }\textbf {\bibinfo {volume}
  {89}},\ \bibinfo {pages} {041004} (\bibinfo {year} {2017})}\BibitemShut
  {NoStop}%
\bibitem [{\citenamefont {Kitaev}(2006)}]{kitaev2006anyons}%
  \BibitemOpen
  \bibfield  {author} {\bibinfo {author} {\bibfnamefont {A.}~\bibnamefont
  {Kitaev}},\ }\bibfield  {title} {\bibinfo {title} {Anyons in an exactly
  solved model and beyond},\ }\href@noop {} {\bibfield  {journal} {\bibinfo
  {journal} {Annals of Physics}\ }\textbf {\bibinfo {volume} {321}},\ \bibinfo
  {pages} {2} (\bibinfo {year} {2006})}\BibitemShut {NoStop}%
\bibitem [{\citenamefont {Nayak}\ \emph {et~al.}(2008)\citenamefont {Nayak},
  \citenamefont {Simon}, \citenamefont {Stern}, \citenamefont {Freedman},\ and\
  \citenamefont {Sarma}}]{nayak2008non}%
  \BibitemOpen
  \bibfield  {author} {\bibinfo {author} {\bibfnamefont {C.}~\bibnamefont
  {Nayak}}, \bibinfo {author} {\bibfnamefont {S.~H.}\ \bibnamefont {Simon}},
  \bibinfo {author} {\bibfnamefont {A.}~\bibnamefont {Stern}}, \bibinfo
  {author} {\bibfnamefont {M.}~\bibnamefont {Freedman}},\ and\ \bibinfo
  {author} {\bibfnamefont {S.~D.}\ \bibnamefont {Sarma}},\ }\bibfield  {title}
  {\bibinfo {title} {Non-abelian anyons and topological quantum computation},\
  }\href@noop {} {\bibfield  {journal} {\bibinfo  {journal} {Reviews of Modern
  Physics}\ }\textbf {\bibinfo {volume} {80}},\ \bibinfo {pages} {1083}
  (\bibinfo {year} {2008})}\BibitemShut {NoStop}%
\bibitem [{\citenamefont {Wilczek}(1990)}]{wilczek1990fractional}%
  \BibitemOpen
  \bibfield  {author} {\bibinfo {author} {\bibfnamefont {F.}~\bibnamefont
  {Wilczek}},\ }\href@noop {} {\emph {\bibinfo {title} {Fractional statistics
  and anyon superconductivity}}},\ Vol.~\bibinfo {volume} {5}\ (\bibinfo
  {publisher} {World scientific},\ \bibinfo {year} {1990})\BibitemShut
  {NoStop}%
\bibitem [{\citenamefont {Stern}(2010)}]{stern2010non}%
  \BibitemOpen
  \bibfield  {author} {\bibinfo {author} {\bibfnamefont {A.}~\bibnamefont
  {Stern}},\ }\bibfield  {title} {\bibinfo {title} {Non-abelian states of
  matter},\ }\href@noop {} {\bibfield  {journal} {\bibinfo  {journal} {Nature}\
  }\textbf {\bibinfo {volume} {464}},\ \bibinfo {pages} {187} (\bibinfo {year}
  {2010})}\BibitemShut {NoStop}%
\bibitem [{\citenamefont {Moore}\ and\ \citenamefont
  {Read}(1991)}]{moore1991nonabelions}%
  \BibitemOpen
  \bibfield  {author} {\bibinfo {author} {\bibfnamefont {G.}~\bibnamefont
  {Moore}}\ and\ \bibinfo {author} {\bibfnamefont {N.}~\bibnamefont {Read}},\
  }\bibfield  {title} {\bibinfo {title} {Nonabelions in the fractional quantum
  hall effect},\ }\href@noop {} {\bibfield  {journal} {\bibinfo  {journal}
  {Nuclear Physics B}\ }\textbf {\bibinfo {volume} {360}},\ \bibinfo {pages}
  {362} (\bibinfo {year} {1991})}\BibitemShut {NoStop}%
\bibitem [{\citenamefont {Wen}(1991)}]{wen1991non}%
  \BibitemOpen
  \bibfield  {author} {\bibinfo {author} {\bibfnamefont {X.-G.}\ \bibnamefont
  {Wen}},\ }\bibfield  {title} {\bibinfo {title} {Non-abelian statistics in the
  fractional quantum hall states},\ }\href@noop {} {\bibfield  {journal}
  {\bibinfo  {journal} {Physical review letters}\ }\textbf {\bibinfo {volume}
  {66}},\ \bibinfo {pages} {802} (\bibinfo {year} {1991})}\BibitemShut
  {NoStop}%
\bibitem [{\citenamefont {Kitaev}(2003)}]{kitaev2003fault}%
  \BibitemOpen
  \bibfield  {author} {\bibinfo {author} {\bibfnamefont {A.~Y.}\ \bibnamefont
  {Kitaev}},\ }\bibfield  {title} {\bibinfo {title} {Fault-tolerant quantum
  computation by anyons},\ }\href@noop {} {\bibfield  {journal} {\bibinfo
  {journal} {Annals of physics}\ }\textbf {\bibinfo {volume} {303}},\ \bibinfo
  {pages} {2} (\bibinfo {year} {2003})}\BibitemShut {NoStop}%
\bibitem [{\citenamefont {Pachos}(2012)}]{pachos2012introduction}%
  \BibitemOpen
  \bibfield  {author} {\bibinfo {author} {\bibfnamefont {J.~K.}\ \bibnamefont
  {Pachos}},\ }\href@noop {} {\emph {\bibinfo {title} {Introduction to
  topological quantum computation}}}\ (\bibinfo  {publisher} {Cambridge
  University Press},\ \bibinfo {year} {2012})\BibitemShut {NoStop}%
\bibitem [{\citenamefont {Stern}\ and\ \citenamefont
  {Lindner}(2013)}]{stern2013topological}%
  \BibitemOpen
  \bibfield  {author} {\bibinfo {author} {\bibfnamefont {A.}~\bibnamefont
  {Stern}}\ and\ \bibinfo {author} {\bibfnamefont {N.~H.}\ \bibnamefont
  {Lindner}},\ }\bibfield  {title} {\bibinfo {title} {Topological quantum
  computation—from basic concepts to first experiments},\ }\href@noop {}
  {\bibfield  {journal} {\bibinfo  {journal} {Science}\ }\textbf {\bibinfo
  {volume} {339}},\ \bibinfo {pages} {1179} (\bibinfo {year}
  {2013})}\BibitemShut {NoStop}%
\bibitem [{\citenamefont {Hermanns}\ \emph {et~al.}(2018)\citenamefont
  {Hermanns}, \citenamefont {Kimchi},\ and\ \citenamefont
  {Knolle}}]{hermanns2018physics}%
  \BibitemOpen
  \bibfield  {author} {\bibinfo {author} {\bibfnamefont {M.}~\bibnamefont
  {Hermanns}}, \bibinfo {author} {\bibfnamefont {I.}~\bibnamefont {Kimchi}},\
  and\ \bibinfo {author} {\bibfnamefont {J.}~\bibnamefont {Knolle}},\
  }\bibfield  {title} {\bibinfo {title} {Physics of the kitaev model:
  fractionalization, dynamic correlations, and material connections},\
  }\href@noop {} {\bibfield  {journal} {\bibinfo  {journal} {Annual Review of
  Condensed Matter Physics}\ }\textbf {\bibinfo {volume} {9}},\ \bibinfo
  {pages} {17} (\bibinfo {year} {2018})}\BibitemShut {NoStop}%
\bibitem [{\citenamefont {Knolle}\ and\ \citenamefont
  {Moessner}(2019)}]{knolle2019field}%
  \BibitemOpen
  \bibfield  {author} {\bibinfo {author} {\bibfnamefont {J.}~\bibnamefont
  {Knolle}}\ and\ \bibinfo {author} {\bibfnamefont {R.}~\bibnamefont
  {Moessner}},\ }\bibfield  {title} {\bibinfo {title} {A field guide to spin
  liquids},\ }\href@noop {} {\bibfield  {journal} {\bibinfo  {journal} {Annual
  Review of Condensed Matter Physics}\ }\textbf {\bibinfo {volume} {10}},\
  \bibinfo {pages} {451} (\bibinfo {year} {2019})}\BibitemShut {NoStop}%
\bibitem [{\citenamefont {Takagi}\ \emph {et~al.}(2019)\citenamefont {Takagi},
  \citenamefont {Takayama}, \citenamefont {Jackeli}, \citenamefont
  {Khaliullin},\ and\ \citenamefont {Nagler}}]{takagi2019concept}%
  \BibitemOpen
  \bibfield  {author} {\bibinfo {author} {\bibfnamefont {H.}~\bibnamefont
  {Takagi}}, \bibinfo {author} {\bibfnamefont {T.}~\bibnamefont {Takayama}},
  \bibinfo {author} {\bibfnamefont {G.}~\bibnamefont {Jackeli}}, \bibinfo
  {author} {\bibfnamefont {G.}~\bibnamefont {Khaliullin}},\ and\ \bibinfo
  {author} {\bibfnamefont {S.~E.}\ \bibnamefont {Nagler}},\ }\bibfield  {title}
  {\bibinfo {title} {Concept and realization of kitaev quantum spin liquids},\
  }\href@noop {} {\bibfield  {journal} {\bibinfo  {journal} {Nature Reviews
  Physics}\ }\textbf {\bibinfo {volume} {1}},\ \bibinfo {pages} {264} (\bibinfo
  {year} {2019})}\BibitemShut {NoStop}%
\bibitem [{\citenamefont {Trebst}\ and\ \citenamefont
  {Hickey}(2022)}]{trebst2022kitaev}%
  \BibitemOpen
  \bibfield  {author} {\bibinfo {author} {\bibfnamefont {S.}~\bibnamefont
  {Trebst}}\ and\ \bibinfo {author} {\bibfnamefont {C.}~\bibnamefont
  {Hickey}},\ }\bibfield  {title} {\bibinfo {title} {Kitaev materials},\
  }\href@noop {} {\bibfield  {journal} {\bibinfo  {journal} {Physics Reports}\
  }\textbf {\bibinfo {volume} {950}},\ \bibinfo {pages} {1} (\bibinfo {year}
  {2022})}\BibitemShut {NoStop}%
\bibitem [{\citenamefont {Banerjee}\ \emph {et~al.}(2017)\citenamefont
  {Banerjee}, \citenamefont {Yan}, \citenamefont {Knolle}, \citenamefont
  {Bridges}, \citenamefont {Stone}, \citenamefont {Lumsden}, \citenamefont
  {Mandrus}, \citenamefont {Tennant}, \citenamefont {Moessner},\ and\
  \citenamefont {Nagler}}]{banerjee2017neutron}%
  \BibitemOpen
  \bibfield  {author} {\bibinfo {author} {\bibfnamefont {A.}~\bibnamefont
  {Banerjee}}, \bibinfo {author} {\bibfnamefont {J.}~\bibnamefont {Yan}},
  \bibinfo {author} {\bibfnamefont {J.}~\bibnamefont {Knolle}}, \bibinfo
  {author} {\bibfnamefont {C.~A.}\ \bibnamefont {Bridges}}, \bibinfo {author}
  {\bibfnamefont {M.~B.}\ \bibnamefont {Stone}}, \bibinfo {author}
  {\bibfnamefont {M.~D.}\ \bibnamefont {Lumsden}}, \bibinfo {author}
  {\bibfnamefont {D.~G.}\ \bibnamefont {Mandrus}}, \bibinfo {author}
  {\bibfnamefont {D.~A.}\ \bibnamefont {Tennant}}, \bibinfo {author}
  {\bibfnamefont {R.}~\bibnamefont {Moessner}},\ and\ \bibinfo {author}
  {\bibfnamefont {S.~E.}\ \bibnamefont {Nagler}},\ }\bibfield  {title}
  {\bibinfo {title} {Neutron scattering in the proximate quantum spin liquid
  $\alpha$-rucl3},\ }\href@noop {} {\bibfield  {journal} {\bibinfo  {journal}
  {Science}\ }\textbf {\bibinfo {volume} {356}},\ \bibinfo {pages} {1055}
  (\bibinfo {year} {2017})}\BibitemShut {NoStop}%
\bibitem [{\citenamefont {Do}\ \emph {et~al.}(2017)\citenamefont {Do},
  \citenamefont {Park}, \citenamefont {Yoshitake}, \citenamefont {Nasu},
  \citenamefont {Motome}, \citenamefont {Kwon}, \citenamefont {Adroja},
  \citenamefont {Voneshen}, \citenamefont {Kim}, \citenamefont {Jang} \emph
  {et~al.}}]{do2017majorana}%
  \BibitemOpen
  \bibfield  {author} {\bibinfo {author} {\bibfnamefont {S.-H.}\ \bibnamefont
  {Do}}, \bibinfo {author} {\bibfnamefont {S.-Y.}\ \bibnamefont {Park}},
  \bibinfo {author} {\bibfnamefont {J.}~\bibnamefont {Yoshitake}}, \bibinfo
  {author} {\bibfnamefont {J.}~\bibnamefont {Nasu}}, \bibinfo {author}
  {\bibfnamefont {Y.}~\bibnamefont {Motome}}, \bibinfo {author} {\bibfnamefont
  {Y.~S.}\ \bibnamefont {Kwon}}, \bibinfo {author} {\bibfnamefont
  {D.}~\bibnamefont {Adroja}}, \bibinfo {author} {\bibfnamefont
  {D.}~\bibnamefont {Voneshen}}, \bibinfo {author} {\bibfnamefont
  {K.}~\bibnamefont {Kim}}, \bibinfo {author} {\bibfnamefont {T.-H.}\
  \bibnamefont {Jang}}, \emph {et~al.},\ }\bibfield  {title} {\bibinfo {title}
  {Majorana fermions in the kitaev quantum spin system $\alpha$-rucl3},\
  }\href@noop {} {\bibfield  {journal} {\bibinfo  {journal} {Nature Physics}\
  }\textbf {\bibinfo {volume} {13}},\ \bibinfo {pages} {1079} (\bibinfo {year}
  {2017})}\BibitemShut {NoStop}%
\bibitem [{\citenamefont {Kasahara}\ \emph {et~al.}(2018)\citenamefont
  {Kasahara}, \citenamefont {Ohnishi}, \citenamefont {Mizukami}, \citenamefont
  {Tanaka}, \citenamefont {Ma}, \citenamefont {Sugii}, \citenamefont {Kurita},
  \citenamefont {Tanaka}, \citenamefont {Nasu}, \citenamefont {Motome} \emph
  {et~al.}}]{kasahara2018majorana}%
  \BibitemOpen
  \bibfield  {author} {\bibinfo {author} {\bibfnamefont {Y.}~\bibnamefont
  {Kasahara}}, \bibinfo {author} {\bibfnamefont {T.}~\bibnamefont {Ohnishi}},
  \bibinfo {author} {\bibfnamefont {Y.}~\bibnamefont {Mizukami}}, \bibinfo
  {author} {\bibfnamefont {O.}~\bibnamefont {Tanaka}}, \bibinfo {author}
  {\bibfnamefont {S.}~\bibnamefont {Ma}}, \bibinfo {author} {\bibfnamefont
  {K.}~\bibnamefont {Sugii}}, \bibinfo {author} {\bibfnamefont
  {N.}~\bibnamefont {Kurita}}, \bibinfo {author} {\bibfnamefont
  {H.}~\bibnamefont {Tanaka}}, \bibinfo {author} {\bibfnamefont
  {J.}~\bibnamefont {Nasu}}, \bibinfo {author} {\bibfnamefont {Y.}~\bibnamefont
  {Motome}}, \emph {et~al.},\ }\bibfield  {title} {\bibinfo {title} {Majorana
  quantization and half-integer thermal quantum hall effect in a kitaev spin
  liquid},\ }\href@noop {} {\bibfield  {journal} {\bibinfo  {journal} {Nature}\
  }\textbf {\bibinfo {volume} {559}},\ \bibinfo {pages} {227} (\bibinfo {year}
  {2018})}\BibitemShut {NoStop}%
\bibitem [{\citenamefont {Yokoi}\ \emph {et~al.}(2021)\citenamefont {Yokoi},
  \citenamefont {Ma}, \citenamefont {Kasahara}, \citenamefont {Kasahara},
  \citenamefont {Shibauchi}, \citenamefont {Kurita}, \citenamefont {Tanaka},
  \citenamefont {Nasu}, \citenamefont {Motome}, \citenamefont {Hickey} \emph
  {et~al.}}]{yokoi2021half}%
  \BibitemOpen
  \bibfield  {author} {\bibinfo {author} {\bibfnamefont {T.}~\bibnamefont
  {Yokoi}}, \bibinfo {author} {\bibfnamefont {S.}~\bibnamefont {Ma}}, \bibinfo
  {author} {\bibfnamefont {Y.}~\bibnamefont {Kasahara}}, \bibinfo {author}
  {\bibfnamefont {S.}~\bibnamefont {Kasahara}}, \bibinfo {author}
  {\bibfnamefont {T.}~\bibnamefont {Shibauchi}}, \bibinfo {author}
  {\bibfnamefont {N.}~\bibnamefont {Kurita}}, \bibinfo {author} {\bibfnamefont
  {H.}~\bibnamefont {Tanaka}}, \bibinfo {author} {\bibfnamefont
  {J.}~\bibnamefont {Nasu}}, \bibinfo {author} {\bibfnamefont {Y.}~\bibnamefont
  {Motome}}, \bibinfo {author} {\bibfnamefont {C.}~\bibnamefont {Hickey}},
  \emph {et~al.},\ }\bibfield  {title} {\bibinfo {title} {Half-integer
  quantized anomalous thermal hall effect in the kitaev material candidate
  $\alpha$-rucl3},\ }\href@noop {} {\bibfield  {journal} {\bibinfo  {journal}
  {Science}\ }\textbf {\bibinfo {volume} {373}},\ \bibinfo {pages} {568}
  (\bibinfo {year} {2021})}\BibitemShut {NoStop}%
\bibitem [{\citenamefont {Bruin}\ \emph {et~al.}(2022)\citenamefont {Bruin},
  \citenamefont {Claus}, \citenamefont {Matsumoto}, \citenamefont {Kurita},
  \citenamefont {Tanaka},\ and\ \citenamefont {Takagi}}]{bruin2022robustness}%
  \BibitemOpen
  \bibfield  {author} {\bibinfo {author} {\bibfnamefont {J.}~\bibnamefont
  {Bruin}}, \bibinfo {author} {\bibfnamefont {R.}~\bibnamefont {Claus}},
  \bibinfo {author} {\bibfnamefont {Y.}~\bibnamefont {Matsumoto}}, \bibinfo
  {author} {\bibfnamefont {N.}~\bibnamefont {Kurita}}, \bibinfo {author}
  {\bibfnamefont {H.}~\bibnamefont {Tanaka}},\ and\ \bibinfo {author}
  {\bibfnamefont {H.}~\bibnamefont {Takagi}},\ }\bibfield  {title} {\bibinfo
  {title} {Robustness of the thermal hall effect close to half-quantization in
  $\alpha$-rucl3},\ }\href@noop {} {\bibfield  {journal} {\bibinfo  {journal}
  {Nature Physics}\ }\textbf {\bibinfo {volume} {18}},\ \bibinfo {pages} {401}
  (\bibinfo {year} {2022})}\BibitemShut {NoStop}%
\bibitem [{\citenamefont {Baek}\ \emph {et~al.}(2017)\citenamefont {Baek},
  \citenamefont {Do}, \citenamefont {Choi}, \citenamefont {Kwon}, \citenamefont
  {Wolter}, \citenamefont {Nishimoto}, \citenamefont {Van Den~Brink},\ and\
  \citenamefont {B{\"u}chner}}]{baek2017evidence}%
  \BibitemOpen
  \bibfield  {author} {\bibinfo {author} {\bibfnamefont {S.-H.}\ \bibnamefont
  {Baek}}, \bibinfo {author} {\bibfnamefont {S.-H.}\ \bibnamefont {Do}},
  \bibinfo {author} {\bibfnamefont {K.-Y.}\ \bibnamefont {Choi}}, \bibinfo
  {author} {\bibfnamefont {Y.~S.}\ \bibnamefont {Kwon}}, \bibinfo {author}
  {\bibfnamefont {A.}~\bibnamefont {Wolter}}, \bibinfo {author} {\bibfnamefont
  {S.}~\bibnamefont {Nishimoto}}, \bibinfo {author} {\bibfnamefont
  {J.}~\bibnamefont {Van Den~Brink}},\ and\ \bibinfo {author} {\bibfnamefont
  {B.}~\bibnamefont {B{\"u}chner}},\ }\bibfield  {title} {\bibinfo {title}
  {Evidence for a field-induced quantum spin liquid in $\alpha$-rucl 3},\
  }\href@noop {} {\bibfield  {journal} {\bibinfo  {journal} {Physical review
  letters}\ }\textbf {\bibinfo {volume} {119}},\ \bibinfo {pages} {037201}
  (\bibinfo {year} {2017})}\BibitemShut {NoStop}%
\bibitem [{\citenamefont {Hentrich}\ \emph {et~al.}(2018)\citenamefont
  {Hentrich}, \citenamefont {Wolter}, \citenamefont {Zotos}, \citenamefont
  {Brenig}, \citenamefont {Nowak}, \citenamefont {Isaeva}, \citenamefont
  {Doert}, \citenamefont {Banerjee}, \citenamefont {Lampen-Kelley},
  \citenamefont {Mandrus} \emph {et~al.}}]{hentrich2018unusual}%
  \BibitemOpen
  \bibfield  {author} {\bibinfo {author} {\bibfnamefont {R.}~\bibnamefont
  {Hentrich}}, \bibinfo {author} {\bibfnamefont {A.~U.}\ \bibnamefont
  {Wolter}}, \bibinfo {author} {\bibfnamefont {X.}~\bibnamefont {Zotos}},
  \bibinfo {author} {\bibfnamefont {W.}~\bibnamefont {Brenig}}, \bibinfo
  {author} {\bibfnamefont {D.}~\bibnamefont {Nowak}}, \bibinfo {author}
  {\bibfnamefont {A.}~\bibnamefont {Isaeva}}, \bibinfo {author} {\bibfnamefont
  {T.}~\bibnamefont {Doert}}, \bibinfo {author} {\bibfnamefont
  {A.}~\bibnamefont {Banerjee}}, \bibinfo {author} {\bibfnamefont
  {P.}~\bibnamefont {Lampen-Kelley}}, \bibinfo {author} {\bibfnamefont {D.~G.}\
  \bibnamefont {Mandrus}}, \emph {et~al.},\ }\bibfield  {title} {\bibinfo
  {title} {Unusual phonon heat transport in $\alpha$- rucl 3: strong
  spin-phonon scattering and field-induced spin gap},\ }\href@noop {}
  {\bibfield  {journal} {\bibinfo  {journal} {Physical review letters}\
  }\textbf {\bibinfo {volume} {120}},\ \bibinfo {pages} {117204} (\bibinfo
  {year} {2018})}\BibitemShut {NoStop}%
\bibitem [{\citenamefont {Bachus}\ \emph {et~al.}(2020)\citenamefont {Bachus},
  \citenamefont {Kaib}, \citenamefont {Tokiwa}, \citenamefont {Jesche},
  \citenamefont {Tsurkan}, \citenamefont {Loidl}, \citenamefont {Winter},
  \citenamefont {Tsirlin}, \citenamefont {Valenti},\ and\ \citenamefont
  {Gegenwart}}]{bachus2020thermodynamic}%
  \BibitemOpen
  \bibfield  {author} {\bibinfo {author} {\bibfnamefont {S.}~\bibnamefont
  {Bachus}}, \bibinfo {author} {\bibfnamefont {D.~A.}\ \bibnamefont {Kaib}},
  \bibinfo {author} {\bibfnamefont {Y.}~\bibnamefont {Tokiwa}}, \bibinfo
  {author} {\bibfnamefont {A.}~\bibnamefont {Jesche}}, \bibinfo {author}
  {\bibfnamefont {V.}~\bibnamefont {Tsurkan}}, \bibinfo {author} {\bibfnamefont
  {A.}~\bibnamefont {Loidl}}, \bibinfo {author} {\bibfnamefont {S.~M.}\
  \bibnamefont {Winter}}, \bibinfo {author} {\bibfnamefont {A.~A.}\
  \bibnamefont {Tsirlin}}, \bibinfo {author} {\bibfnamefont {R.}~\bibnamefont
  {Valenti}},\ and\ \bibinfo {author} {\bibfnamefont {P.}~\bibnamefont
  {Gegenwart}},\ }\bibfield  {title} {\bibinfo {title} {Thermodynamic
  perspective on field-induced behavior of $\alpha$- rucl 3},\ }\href@noop {}
  {\bibfield  {journal} {\bibinfo  {journal} {Physical Review Letters}\
  }\textbf {\bibinfo {volume} {125}},\ \bibinfo {pages} {097203} (\bibinfo
  {year} {2020})}\BibitemShut {NoStop}%
\bibitem [{\citenamefont {Leahy}\ \emph {et~al.}(2017)\citenamefont {Leahy},
  \citenamefont {Pocs}, \citenamefont {Siegfried}, \citenamefont {Graf},
  \citenamefont {Do}, \citenamefont {Choi}, \citenamefont {Normand},\ and\
  \citenamefont {Lee}}]{leahy2017anomalous}%
  \BibitemOpen
  \bibfield  {author} {\bibinfo {author} {\bibfnamefont {I.~A.}\ \bibnamefont
  {Leahy}}, \bibinfo {author} {\bibfnamefont {C.~A.}\ \bibnamefont {Pocs}},
  \bibinfo {author} {\bibfnamefont {P.~E.}\ \bibnamefont {Siegfried}}, \bibinfo
  {author} {\bibfnamefont {D.}~\bibnamefont {Graf}}, \bibinfo {author}
  {\bibfnamefont {S.-H.}\ \bibnamefont {Do}}, \bibinfo {author} {\bibfnamefont
  {K.-Y.}\ \bibnamefont {Choi}}, \bibinfo {author} {\bibfnamefont
  {B.}~\bibnamefont {Normand}},\ and\ \bibinfo {author} {\bibfnamefont
  {M.}~\bibnamefont {Lee}},\ }\bibfield  {title} {\bibinfo {title} {Anomalous
  thermal conductivity and magnetic torque response in the honeycomb magnet
  $\alpha$- rucl 3},\ }\href@noop {} {\bibfield  {journal} {\bibinfo  {journal}
  {Physical review letters}\ }\textbf {\bibinfo {volume} {118}},\ \bibinfo
  {pages} {187203} (\bibinfo {year} {2017})}\BibitemShut {NoStop}%
\bibitem [{\citenamefont {Jan{\v{s}}a}\ \emph {et~al.}(2018)\citenamefont
  {Jan{\v{s}}a}, \citenamefont {Zorko}, \citenamefont {Gomil{\v{s}}ek},
  \citenamefont {Pregelj}, \citenamefont {Kr{\"a}mer}, \citenamefont {Biner},
  \citenamefont {Biffin}, \citenamefont {R{\"u}egg},\ and\ \citenamefont
  {Klanj{\v{s}}ek}}]{janvsa2018observation}%
  \BibitemOpen
  \bibfield  {author} {\bibinfo {author} {\bibfnamefont {N.}~\bibnamefont
  {Jan{\v{s}}a}}, \bibinfo {author} {\bibfnamefont {A.}~\bibnamefont {Zorko}},
  \bibinfo {author} {\bibfnamefont {M.}~\bibnamefont {Gomil{\v{s}}ek}},
  \bibinfo {author} {\bibfnamefont {M.}~\bibnamefont {Pregelj}}, \bibinfo
  {author} {\bibfnamefont {K.~W.}\ \bibnamefont {Kr{\"a}mer}}, \bibinfo
  {author} {\bibfnamefont {D.}~\bibnamefont {Biner}}, \bibinfo {author}
  {\bibfnamefont {A.}~\bibnamefont {Biffin}}, \bibinfo {author} {\bibfnamefont
  {C.}~\bibnamefont {R{\"u}egg}},\ and\ \bibinfo {author} {\bibfnamefont
  {M.}~\bibnamefont {Klanj{\v{s}}ek}},\ }\bibfield  {title} {\bibinfo {title}
  {Observation of two types of fractional excitation in the kitaev honeycomb
  magnet},\ }\href@noop {} {\bibfield  {journal} {\bibinfo  {journal} {Nature
  physics}\ }\textbf {\bibinfo {volume} {14}},\ \bibinfo {pages} {786}
  (\bibinfo {year} {2018})}\BibitemShut {NoStop}%
\bibitem [{\citenamefont {Winter}\ \emph {et~al.}(2018)\citenamefont {Winter},
  \citenamefont {Riedl}, \citenamefont {Kaib}, \citenamefont {Coldea},\ and\
  \citenamefont {Valent{\'\i}}}]{winter2018probing}%
  \BibitemOpen
  \bibfield  {author} {\bibinfo {author} {\bibfnamefont {S.~M.}\ \bibnamefont
  {Winter}}, \bibinfo {author} {\bibfnamefont {K.}~\bibnamefont {Riedl}},
  \bibinfo {author} {\bibfnamefont {D.}~\bibnamefont {Kaib}}, \bibinfo {author}
  {\bibfnamefont {R.}~\bibnamefont {Coldea}},\ and\ \bibinfo {author}
  {\bibfnamefont {R.}~\bibnamefont {Valent{\'\i}}},\ }\bibfield  {title}
  {\bibinfo {title} {Probing $\alpha$- rucl 3 beyond magnetic order: Effects of
  temperature and magnetic field},\ }\href@noop {} {\bibfield  {journal}
  {\bibinfo  {journal} {Physical review letters}\ }\textbf {\bibinfo {volume}
  {120}},\ \bibinfo {pages} {077203} (\bibinfo {year} {2018})}\BibitemShut
  {NoStop}%
\bibitem [{\citenamefont {Chern}\ \emph {et~al.}(2021)\citenamefont {Chern},
  \citenamefont {Zhang},\ and\ \citenamefont {Kim}}]{chern2021sign}%
  \BibitemOpen
  \bibfield  {author} {\bibinfo {author} {\bibfnamefont {L.~E.}\ \bibnamefont
  {Chern}}, \bibinfo {author} {\bibfnamefont {E.~Z.}\ \bibnamefont {Zhang}},\
  and\ \bibinfo {author} {\bibfnamefont {Y.~B.}\ \bibnamefont {Kim}},\
  }\bibfield  {title} {\bibinfo {title} {Sign structure of thermal hall
  conductivity and topological magnons for in-plane field polarized kitaev
  magnets},\ }\href@noop {} {\bibfield  {journal} {\bibinfo  {journal}
  {Physical Review Letters}\ }\textbf {\bibinfo {volume} {126}},\ \bibinfo
  {pages} {147201} (\bibinfo {year} {2021})}\BibitemShut {NoStop}%
\bibitem [{\citenamefont {Georgescu}\ \emph {et~al.}(2014)\citenamefont
  {Georgescu}, \citenamefont {Ashhab},\ and\ \citenamefont
  {Nori}}]{georgescu2014quantum}%
  \BibitemOpen
  \bibfield  {author} {\bibinfo {author} {\bibfnamefont {I.~M.}\ \bibnamefont
  {Georgescu}}, \bibinfo {author} {\bibfnamefont {S.}~\bibnamefont {Ashhab}},\
  and\ \bibinfo {author} {\bibfnamefont {F.}~\bibnamefont {Nori}},\ }\bibfield
  {title} {\bibinfo {title} {Quantum simulation},\ }\href@noop {} {\bibfield
  {journal} {\bibinfo  {journal} {Reviews of Modern Physics}\ }\textbf
  {\bibinfo {volume} {86}},\ \bibinfo {pages} {153} (\bibinfo {year}
  {2014})}\BibitemShut {NoStop}%
\bibitem [{\citenamefont {Daley}\ \emph {et~al.}(2022)\citenamefont {Daley},
  \citenamefont {Bloch}, \citenamefont {Kokail}, \citenamefont {Flannigan},
  \citenamefont {Pearson}, \citenamefont {Troyer},\ and\ \citenamefont
  {Zoller}}]{daley2022practical}%
  \BibitemOpen
  \bibfield  {author} {\bibinfo {author} {\bibfnamefont {A.~J.}\ \bibnamefont
  {Daley}}, \bibinfo {author} {\bibfnamefont {I.}~\bibnamefont {Bloch}},
  \bibinfo {author} {\bibfnamefont {C.}~\bibnamefont {Kokail}}, \bibinfo
  {author} {\bibfnamefont {S.}~\bibnamefont {Flannigan}}, \bibinfo {author}
  {\bibfnamefont {N.}~\bibnamefont {Pearson}}, \bibinfo {author} {\bibfnamefont
  {M.}~\bibnamefont {Troyer}},\ and\ \bibinfo {author} {\bibfnamefont
  {P.}~\bibnamefont {Zoller}},\ }\bibfield  {title} {\bibinfo {title}
  {Practical quantum advantage in quantum simulation},\ }\href@noop {}
  {\bibfield  {journal} {\bibinfo  {journal} {Nature}\ }\textbf {\bibinfo
  {volume} {607}},\ \bibinfo {pages} {667} (\bibinfo {year}
  {2022})}\BibitemShut {NoStop}%
\bibitem [{\citenamefont {Duan}\ \emph {et~al.}(2003)\citenamefont {Duan},
  \citenamefont {Demler},\ and\ \citenamefont {Lukin}}]{duan2003controlling}%
  \BibitemOpen
  \bibfield  {author} {\bibinfo {author} {\bibfnamefont {L.-M.}\ \bibnamefont
  {Duan}}, \bibinfo {author} {\bibfnamefont {E.}~\bibnamefont {Demler}},\ and\
  \bibinfo {author} {\bibfnamefont {M.~D.}\ \bibnamefont {Lukin}},\ }\bibfield
  {title} {\bibinfo {title} {Controlling spin exchange interactions of
  ultracold atoms in optical lattices},\ }\href@noop {} {\bibfield  {journal}
  {\bibinfo  {journal} {Physical review letters}\ }\textbf {\bibinfo {volume}
  {91}},\ \bibinfo {pages} {090402} (\bibinfo {year} {2003})}\BibitemShut
  {NoStop}%
\bibitem [{\citenamefont {Bookatz}\ \emph {et~al.}(2014)\citenamefont
  {Bookatz}, \citenamefont {Wocjan},\ and\ \citenamefont
  {Viola}}]{bookatz2014hamiltonian}%
  \BibitemOpen
  \bibfield  {author} {\bibinfo {author} {\bibfnamefont {A.~D.}\ \bibnamefont
  {Bookatz}}, \bibinfo {author} {\bibfnamefont {P.}~\bibnamefont {Wocjan}},\
  and\ \bibinfo {author} {\bibfnamefont {L.}~\bibnamefont {Viola}},\ }\bibfield
   {title} {\bibinfo {title} {Hamiltonian quantum simulation with
  bounded-strength controls},\ }\href@noop {} {\bibfield  {journal} {\bibinfo
  {journal} {New Journal of Physics}\ }\textbf {\bibinfo {volume} {16}},\
  \bibinfo {pages} {045021} (\bibinfo {year} {2014})}\BibitemShut {NoStop}%
\bibitem [{\citenamefont {Po}\ \emph {et~al.}(2017)\citenamefont {Po},
  \citenamefont {Fidkowski}, \citenamefont {Vishwanath},\ and\ \citenamefont
  {Potter}}]{po2017radical}%
  \BibitemOpen
  \bibfield  {author} {\bibinfo {author} {\bibfnamefont {H.~C.}\ \bibnamefont
  {Po}}, \bibinfo {author} {\bibfnamefont {L.}~\bibnamefont {Fidkowski}},
  \bibinfo {author} {\bibfnamefont {A.}~\bibnamefont {Vishwanath}},\ and\
  \bibinfo {author} {\bibfnamefont {A.~C.}\ \bibnamefont {Potter}},\ }\bibfield
   {title} {\bibinfo {title} {Radical chiral floquet phases in a periodically
  driven kitaev model and beyond},\ }\href@noop {} {\bibfield  {journal}
  {\bibinfo  {journal} {Physical Review B}\ }\textbf {\bibinfo {volume} {96}},\
  \bibinfo {pages} {245116} (\bibinfo {year} {2017})}\BibitemShut {NoStop}%
\bibitem [{\citenamefont {Kalinowski}\ \emph {et~al.}(2022)\citenamefont
  {Kalinowski}, \citenamefont {Maskara},\ and\ \citenamefont
  {Lukin}}]{lukin2022non}%
  \BibitemOpen
  \bibfield  {author} {\bibinfo {author} {\bibfnamefont {M.}~\bibnamefont
  {Kalinowski}}, \bibinfo {author} {\bibfnamefont {N.}~\bibnamefont
  {Maskara}},\ and\ \bibinfo {author} {\bibfnamefont {M.~D.}\ \bibnamefont
  {Lukin}},\ }\bibfield  {title} {\bibinfo {title} {Non-abelian floquet spin
  liquids in a digital rydberg simulator},\ }\href@noop {} {\bibfield
  {journal} {\bibinfo  {journal} {arXiv preprint arXiv:2211.00017}\ } (\bibinfo
  {year} {2022})}\BibitemShut {NoStop}%
\bibitem [{\citenamefont {Sun}\ \emph {et~al.}(2023)\citenamefont {Sun},
  \citenamefont {Goldman}, \citenamefont {Aidelsburger},\ and\ \citenamefont
  {Bukov}}]{Monika020329}%
  \BibitemOpen
  \bibfield  {author} {\bibinfo {author} {\bibfnamefont {B.-Y.}\ \bibnamefont
  {Sun}}, \bibinfo {author} {\bibfnamefont {N.}~\bibnamefont {Goldman}},
  \bibinfo {author} {\bibfnamefont {M.}~\bibnamefont {Aidelsburger}},\ and\
  \bibinfo {author} {\bibfnamefont {M.}~\bibnamefont {Bukov}},\ }\bibfield
  {title} {\bibinfo {title} {Engineering and probing non-abelian chiral spin
  liquids using periodically driven ultracold atoms},\ }\href@noop {}
  {\bibfield  {journal} {\bibinfo  {journal} {PRX Quantum}\ }\textbf {\bibinfo
  {volume} {4}},\ \bibinfo {pages} {020329} (\bibinfo {year}
  {2023})}\BibitemShut {NoStop}%
\bibitem [{\citenamefont {Browaeys}\ and\ \citenamefont
  {Lahaye}(2020)}]{browaeys2020many}%
  \BibitemOpen
  \bibfield  {author} {\bibinfo {author} {\bibfnamefont {A.}~\bibnamefont
  {Browaeys}}\ and\ \bibinfo {author} {\bibfnamefont {T.}~\bibnamefont
  {Lahaye}},\ }\bibfield  {title} {\bibinfo {title} {Many-body physics with
  individually controlled rydberg atoms},\ }\href@noop {} {\bibfield  {journal}
  {\bibinfo  {journal} {Nature Physics}\ }\textbf {\bibinfo {volume} {16}},\
  \bibinfo {pages} {132} (\bibinfo {year} {2020})}\BibitemShut {NoStop}%
\bibitem [{\citenamefont {Glaetzle}\ \emph {et~al.}(2014)\citenamefont
  {Glaetzle}, \citenamefont {Dalmonte}, \citenamefont {Nath}, \citenamefont
  {Rousochatzakis}, \citenamefont {Moessner},\ and\ \citenamefont
  {Zoller}}]{glaetzle2014quantum}%
  \BibitemOpen
  \bibfield  {author} {\bibinfo {author} {\bibfnamefont {A.~W.}\ \bibnamefont
  {Glaetzle}}, \bibinfo {author} {\bibfnamefont {M.}~\bibnamefont {Dalmonte}},
  \bibinfo {author} {\bibfnamefont {R.}~\bibnamefont {Nath}}, \bibinfo {author}
  {\bibfnamefont {I.}~\bibnamefont {Rousochatzakis}}, \bibinfo {author}
  {\bibfnamefont {R.}~\bibnamefont {Moessner}},\ and\ \bibinfo {author}
  {\bibfnamefont {P.}~\bibnamefont {Zoller}},\ }\bibfield  {title} {\bibinfo
  {title} {Quantum spin-ice and dimer models with rydberg atoms},\ }\href@noop
  {} {\bibfield  {journal} {\bibinfo  {journal} {Physical Review X}\ }\textbf
  {\bibinfo {volume} {4}},\ \bibinfo {pages} {041037} (\bibinfo {year}
  {2014})}\BibitemShut {NoStop}%
\bibitem [{\citenamefont {Nguyen}\ \emph {et~al.}(2018)\citenamefont {Nguyen},
  \citenamefont {Raimond}, \citenamefont {Sayrin}, \citenamefont {Cortinas},
  \citenamefont {Cantat-Moltrecht}, \citenamefont {Assemat}, \citenamefont
  {Dotsenko}, \citenamefont {Gleyzes}, \citenamefont {Haroche}, \citenamefont
  {Roux} \emph {et~al.}}]{nguyen2018towards}%
  \BibitemOpen
  \bibfield  {author} {\bibinfo {author} {\bibfnamefont {T.~L.}\ \bibnamefont
  {Nguyen}}, \bibinfo {author} {\bibfnamefont {J.-M.}\ \bibnamefont {Raimond}},
  \bibinfo {author} {\bibfnamefont {C.}~\bibnamefont {Sayrin}}, \bibinfo
  {author} {\bibfnamefont {R.}~\bibnamefont {Cortinas}}, \bibinfo {author}
  {\bibfnamefont {T.}~\bibnamefont {Cantat-Moltrecht}}, \bibinfo {author}
  {\bibfnamefont {F.}~\bibnamefont {Assemat}}, \bibinfo {author} {\bibfnamefont
  {I.}~\bibnamefont {Dotsenko}}, \bibinfo {author} {\bibfnamefont
  {S.}~\bibnamefont {Gleyzes}}, \bibinfo {author} {\bibfnamefont
  {S.}~\bibnamefont {Haroche}}, \bibinfo {author} {\bibfnamefont
  {G.}~\bibnamefont {Roux}}, \emph {et~al.},\ }\bibfield  {title} {\bibinfo
  {title} {Towards quantum simulation with circular rydberg atoms},\
  }\href@noop {} {\bibfield  {journal} {\bibinfo  {journal} {Physical Review
  X}\ }\textbf {\bibinfo {volume} {8}},\ \bibinfo {pages} {011032} (\bibinfo
  {year} {2018})}\BibitemShut {NoStop}%
\bibitem [{\citenamefont {Celi}\ \emph {et~al.}(2020)\citenamefont {Celi},
  \citenamefont {Vermersch}, \citenamefont {Viyuela}, \citenamefont {Pichler},
  \citenamefont {Lukin},\ and\ \citenamefont {Zoller}}]{celi2020emerging}%
  \BibitemOpen
  \bibfield  {author} {\bibinfo {author} {\bibfnamefont {A.}~\bibnamefont
  {Celi}}, \bibinfo {author} {\bibfnamefont {B.}~\bibnamefont {Vermersch}},
  \bibinfo {author} {\bibfnamefont {O.}~\bibnamefont {Viyuela}}, \bibinfo
  {author} {\bibfnamefont {H.}~\bibnamefont {Pichler}}, \bibinfo {author}
  {\bibfnamefont {M.~D.}\ \bibnamefont {Lukin}},\ and\ \bibinfo {author}
  {\bibfnamefont {P.}~\bibnamefont {Zoller}},\ }\bibfield  {title} {\bibinfo
  {title} {Emerging two-dimensional gauge theories in rydberg configurable
  arrays},\ }\href@noop {} {\bibfield  {journal} {\bibinfo  {journal} {Physical
  Review X}\ }\textbf {\bibinfo {volume} {10}},\ \bibinfo {pages} {021057}
  (\bibinfo {year} {2020})}\BibitemShut {NoStop}%
\bibitem [{\citenamefont {Mazza}\ \emph {et~al.}(2020)\citenamefont {Mazza},
  \citenamefont {Schmidt},\ and\ \citenamefont
  {Lesanovsky}}]{mazza2020vibrational}%
  \BibitemOpen
  \bibfield  {author} {\bibinfo {author} {\bibfnamefont {P.~P.}\ \bibnamefont
  {Mazza}}, \bibinfo {author} {\bibfnamefont {R.}~\bibnamefont {Schmidt}},\
  and\ \bibinfo {author} {\bibfnamefont {I.}~\bibnamefont {Lesanovsky}},\
  }\bibfield  {title} {\bibinfo {title} {Vibrational dressing in kinetically
  constrained rydberg spin systems},\ }\href@noop {} {\bibfield  {journal}
  {\bibinfo  {journal} {Physical Review Letters}\ }\textbf {\bibinfo {volume}
  {125}},\ \bibinfo {pages} {033602} (\bibinfo {year} {2020})}\BibitemShut
  {NoStop}%
\bibitem [{\citenamefont {Lee}\ \emph {et~al.}(2023)\citenamefont {Lee},
  \citenamefont {Ramette}, \citenamefont {Metlitski}, \citenamefont
  {Vuleti{\'c}}, \citenamefont {Ho},\ and\ \citenamefont
  {Choi}}]{lee2023landau}%
  \BibitemOpen
  \bibfield  {author} {\bibinfo {author} {\bibfnamefont {J.~Y.}\ \bibnamefont
  {Lee}}, \bibinfo {author} {\bibfnamefont {J.}~\bibnamefont {Ramette}},
  \bibinfo {author} {\bibfnamefont {M.~A.}\ \bibnamefont {Metlitski}}, \bibinfo
  {author} {\bibfnamefont {V.}~\bibnamefont {Vuleti{\'c}}}, \bibinfo {author}
  {\bibfnamefont {W.~W.}\ \bibnamefont {Ho}},\ and\ \bibinfo {author}
  {\bibfnamefont {S.}~\bibnamefont {Choi}},\ }\bibfield  {title} {\bibinfo
  {title} {Landau-forbidden quantum criticality in rydberg quantum
  simulators},\ }\href@noop {} {\bibfield  {journal} {\bibinfo  {journal}
  {Physical Review Letters}\ }\textbf {\bibinfo {volume} {131}},\ \bibinfo
  {pages} {083601} (\bibinfo {year} {2023})}\BibitemShut {NoStop}%
\bibitem [{\citenamefont {Nishad}\ \emph {et~al.}(2023)\citenamefont {Nishad},
  \citenamefont {Keselman}, \citenamefont {Lahaye}, \citenamefont {Browaeys},\
  and\ \citenamefont {Tsesses}}]{nishad2023quantum}%
  \BibitemOpen
  \bibfield  {author} {\bibinfo {author} {\bibfnamefont {N.}~\bibnamefont
  {Nishad}}, \bibinfo {author} {\bibfnamefont {A.}~\bibnamefont {Keselman}},
  \bibinfo {author} {\bibfnamefont {T.}~\bibnamefont {Lahaye}}, \bibinfo
  {author} {\bibfnamefont {A.}~\bibnamefont {Browaeys}},\ and\ \bibinfo
  {author} {\bibfnamefont {S.}~\bibnamefont {Tsesses}},\ }\bibfield  {title}
  {\bibinfo {title} {Quantum simulation of generic spin exchange models in
  floquet-engineered rydberg atom arrays},\ }\href@noop {} {\bibfield
  {journal} {\bibinfo  {journal} {arXiv preprint arXiv:2306.07041}\ } (\bibinfo
  {year} {2023})}\BibitemShut {NoStop}%
\bibitem [{\citenamefont {Kunimi}\ \emph {et~al.}(2023)\citenamefont {Kunimi},
  \citenamefont {Tomita}, \citenamefont {Katsura},\ and\ \citenamefont
  {Kato}}]{kunimi2023proposal}%
  \BibitemOpen
  \bibfield  {author} {\bibinfo {author} {\bibfnamefont {M.}~\bibnamefont
  {Kunimi}}, \bibinfo {author} {\bibfnamefont {T.}~\bibnamefont {Tomita}},
  \bibinfo {author} {\bibfnamefont {H.}~\bibnamefont {Katsura}},\ and\ \bibinfo
  {author} {\bibfnamefont {Y.}~\bibnamefont {Kato}},\ }\bibfield  {title}
  {\bibinfo {title} {Proposal for realizing quantum spin models with
  dzyaloshinskii-moriya interaction using rydberg atoms},\ }\href@noop {}
  {\bibfield  {journal} {\bibinfo  {journal} {arXiv preprint arXiv:2306.05591}\
  } (\bibinfo {year} {2023})}\BibitemShut {NoStop}%
\bibitem [{\citenamefont {Barredo}\ \emph {et~al.}(2016)\citenamefont
  {Barredo}, \citenamefont {De~L{\'e}s{\'e}leuc}, \citenamefont {Lienhard},
  \citenamefont {Lahaye},\ and\ \citenamefont {Browaeys}}]{barredo2016atom}%
  \BibitemOpen
  \bibfield  {author} {\bibinfo {author} {\bibfnamefont {D.}~\bibnamefont
  {Barredo}}, \bibinfo {author} {\bibfnamefont {S.}~\bibnamefont
  {De~L{\'e}s{\'e}leuc}}, \bibinfo {author} {\bibfnamefont {V.}~\bibnamefont
  {Lienhard}}, \bibinfo {author} {\bibfnamefont {T.}~\bibnamefont {Lahaye}},\
  and\ \bibinfo {author} {\bibfnamefont {A.}~\bibnamefont {Browaeys}},\
  }\bibfield  {title} {\bibinfo {title} {An atom-by-atom assembler of
  defect-free arbitrary two-dimensional atomic arrays},\ }\href@noop {}
  {\bibfield  {journal} {\bibinfo  {journal} {Science}\ }\textbf {\bibinfo
  {volume} {354}},\ \bibinfo {pages} {1021} (\bibinfo {year}
  {2016})}\BibitemShut {NoStop}%
\bibitem [{\citenamefont {Endres}\ \emph {et~al.}(2016)\citenamefont {Endres},
  \citenamefont {Bernien}, \citenamefont {Keesling}, \citenamefont {Levine},
  \citenamefont {Anschuetz}, \citenamefont {Krajenbrink}, \citenamefont
  {Senko}, \citenamefont {Vuletic}, \citenamefont {Greiner},\ and\
  \citenamefont {Lukin}}]{endres2016atom}%
  \BibitemOpen
  \bibfield  {author} {\bibinfo {author} {\bibfnamefont {M.}~\bibnamefont
  {Endres}}, \bibinfo {author} {\bibfnamefont {H.}~\bibnamefont {Bernien}},
  \bibinfo {author} {\bibfnamefont {A.}~\bibnamefont {Keesling}}, \bibinfo
  {author} {\bibfnamefont {H.}~\bibnamefont {Levine}}, \bibinfo {author}
  {\bibfnamefont {E.~R.}\ \bibnamefont {Anschuetz}}, \bibinfo {author}
  {\bibfnamefont {A.}~\bibnamefont {Krajenbrink}}, \bibinfo {author}
  {\bibfnamefont {C.}~\bibnamefont {Senko}}, \bibinfo {author} {\bibfnamefont
  {V.}~\bibnamefont {Vuletic}}, \bibinfo {author} {\bibfnamefont
  {M.}~\bibnamefont {Greiner}},\ and\ \bibinfo {author} {\bibfnamefont {M.~D.}\
  \bibnamefont {Lukin}},\ }\bibfield  {title} {\bibinfo {title} {Atom-by-atom
  assembly of defect-free one-dimensional cold atom arrays},\ }\href@noop {}
  {\bibfield  {journal} {\bibinfo  {journal} {Science}\ }\textbf {\bibinfo
  {volume} {354}},\ \bibinfo {pages} {1024} (\bibinfo {year}
  {2016})}\BibitemShut {NoStop}%
\bibitem [{\citenamefont {Kim}\ \emph {et~al.}(2016)\citenamefont {Kim},
  \citenamefont {Lee}, \citenamefont {Lee}, \citenamefont {Jo}, \citenamefont
  {Song},\ and\ \citenamefont {Ahn}}]{kim2016situ}%
  \BibitemOpen
  \bibfield  {author} {\bibinfo {author} {\bibfnamefont {H.}~\bibnamefont
  {Kim}}, \bibinfo {author} {\bibfnamefont {W.}~\bibnamefont {Lee}}, \bibinfo
  {author} {\bibfnamefont {H.-g.}\ \bibnamefont {Lee}}, \bibinfo {author}
  {\bibfnamefont {H.}~\bibnamefont {Jo}}, \bibinfo {author} {\bibfnamefont
  {Y.}~\bibnamefont {Song}},\ and\ \bibinfo {author} {\bibfnamefont
  {J.}~\bibnamefont {Ahn}},\ }\bibfield  {title} {\bibinfo {title} {In situ
  single-atom array synthesis using dynamic holographic optical tweezers},\
  }\href@noop {} {\bibfield  {journal} {\bibinfo  {journal} {Nature
  communications}\ }\textbf {\bibinfo {volume} {7}},\ \bibinfo {pages} {13317}
  (\bibinfo {year} {2016})}\BibitemShut {NoStop}%
\bibitem [{\citenamefont {Barredo}\ \emph {et~al.}(2018)\citenamefont
  {Barredo}, \citenamefont {Lienhard}, \citenamefont {De~Leseleuc},
  \citenamefont {Lahaye},\ and\ \citenamefont
  {Browaeys}}]{barredo2018synthetic}%
  \BibitemOpen
  \bibfield  {author} {\bibinfo {author} {\bibfnamefont {D.}~\bibnamefont
  {Barredo}}, \bibinfo {author} {\bibfnamefont {V.}~\bibnamefont {Lienhard}},
  \bibinfo {author} {\bibfnamefont {S.}~\bibnamefont {De~Leseleuc}}, \bibinfo
  {author} {\bibfnamefont {T.}~\bibnamefont {Lahaye}},\ and\ \bibinfo {author}
  {\bibfnamefont {A.}~\bibnamefont {Browaeys}},\ }\bibfield  {title} {\bibinfo
  {title} {Synthetic three-dimensional atomic structures assembled atom by
  atom},\ }\href@noop {} {\bibfield  {journal} {\bibinfo  {journal} {Nature}\
  }\textbf {\bibinfo {volume} {561}},\ \bibinfo {pages} {79} (\bibinfo {year}
  {2018})}\BibitemShut {NoStop}%
\bibitem [{\citenamefont {De~Mello}\ \emph {et~al.}(2019)\citenamefont
  {De~Mello}, \citenamefont {Sch{\"a}ffner}, \citenamefont {Werkmann},
  \citenamefont {Preuschoff}, \citenamefont {Kohfahl}, \citenamefont
  {Schlosser},\ and\ \citenamefont {Birkl}}]{de2019defect}%
  \BibitemOpen
  \bibfield  {author} {\bibinfo {author} {\bibfnamefont {D.~O.}\ \bibnamefont
  {De~Mello}}, \bibinfo {author} {\bibfnamefont {D.}~\bibnamefont
  {Sch{\"a}ffner}}, \bibinfo {author} {\bibfnamefont {J.}~\bibnamefont
  {Werkmann}}, \bibinfo {author} {\bibfnamefont {T.}~\bibnamefont
  {Preuschoff}}, \bibinfo {author} {\bibfnamefont {L.}~\bibnamefont {Kohfahl}},
  \bibinfo {author} {\bibfnamefont {M.}~\bibnamefont {Schlosser}},\ and\
  \bibinfo {author} {\bibfnamefont {G.}~\bibnamefont {Birkl}},\ }\bibfield
  {title} {\bibinfo {title} {Defect-free assembly of 2d clusters of more than
  100 single-atom quantum systems},\ }\href@noop {} {\bibfield  {journal}
  {\bibinfo  {journal} {Physical review letters}\ }\textbf {\bibinfo {volume}
  {122}},\ \bibinfo {pages} {203601} (\bibinfo {year} {2019})}\BibitemShut
  {NoStop}%
\bibitem [{\citenamefont {Saffman}\ \emph {et~al.}(2010)\citenamefont
  {Saffman}, \citenamefont {Walker},\ and\ \citenamefont
  {M{\o}lmer}}]{saffman2010quantum}%
  \BibitemOpen
  \bibfield  {author} {\bibinfo {author} {\bibfnamefont {M.}~\bibnamefont
  {Saffman}}, \bibinfo {author} {\bibfnamefont {T.~G.}\ \bibnamefont
  {Walker}},\ and\ \bibinfo {author} {\bibfnamefont {K.}~\bibnamefont
  {M{\o}lmer}},\ }\bibfield  {title} {\bibinfo {title} {Quantum information
  with rydberg atoms},\ }\href@noop {} {\bibfield  {journal} {\bibinfo
  {journal} {Reviews of modern physics}\ }\textbf {\bibinfo {volume} {82}},\
  \bibinfo {pages} {2313} (\bibinfo {year} {2010})}\BibitemShut {NoStop}%
\bibitem [{\citenamefont {Labuhn}\ \emph {et~al.}(2016)\citenamefont {Labuhn},
  \citenamefont {Barredo}, \citenamefont {Ravets}, \citenamefont
  {De~L{\'e}s{\'e}leuc}, \citenamefont {Macr{\`\i}}, \citenamefont {Lahaye},\
  and\ \citenamefont {Browaeys}}]{labuhn2016tunable}%
  \BibitemOpen
  \bibfield  {author} {\bibinfo {author} {\bibfnamefont {H.}~\bibnamefont
  {Labuhn}}, \bibinfo {author} {\bibfnamefont {D.}~\bibnamefont {Barredo}},
  \bibinfo {author} {\bibfnamefont {S.}~\bibnamefont {Ravets}}, \bibinfo
  {author} {\bibfnamefont {S.}~\bibnamefont {De~L{\'e}s{\'e}leuc}}, \bibinfo
  {author} {\bibfnamefont {T.}~\bibnamefont {Macr{\`\i}}}, \bibinfo {author}
  {\bibfnamefont {T.}~\bibnamefont {Lahaye}},\ and\ \bibinfo {author}
  {\bibfnamefont {A.}~\bibnamefont {Browaeys}},\ }\bibfield  {title} {\bibinfo
  {title} {Tunable two-dimensional arrays of single rydberg atoms for realizing
  quantum ising models},\ }\href@noop {} {\bibfield  {journal} {\bibinfo
  {journal} {Nature}\ }\textbf {\bibinfo {volume} {534}},\ \bibinfo {pages}
  {667} (\bibinfo {year} {2016})}\BibitemShut {NoStop}%
\bibitem [{\citenamefont {Bernien}\ \emph {et~al.}(2017)\citenamefont
  {Bernien}, \citenamefont {Schwartz}, \citenamefont {Keesling}, \citenamefont
  {Levine}, \citenamefont {Omran}, \citenamefont {Pichler}, \citenamefont
  {Choi}, \citenamefont {Zibrov}, \citenamefont {Endres}, \citenamefont
  {Greiner} \emph {et~al.}}]{bernien2017probing}%
  \BibitemOpen
  \bibfield  {author} {\bibinfo {author} {\bibfnamefont {H.}~\bibnamefont
  {Bernien}}, \bibinfo {author} {\bibfnamefont {S.}~\bibnamefont {Schwartz}},
  \bibinfo {author} {\bibfnamefont {A.}~\bibnamefont {Keesling}}, \bibinfo
  {author} {\bibfnamefont {H.}~\bibnamefont {Levine}}, \bibinfo {author}
  {\bibfnamefont {A.}~\bibnamefont {Omran}}, \bibinfo {author} {\bibfnamefont
  {H.}~\bibnamefont {Pichler}}, \bibinfo {author} {\bibfnamefont
  {S.}~\bibnamefont {Choi}}, \bibinfo {author} {\bibfnamefont {A.~S.}\
  \bibnamefont {Zibrov}}, \bibinfo {author} {\bibfnamefont {M.}~\bibnamefont
  {Endres}}, \bibinfo {author} {\bibfnamefont {M.}~\bibnamefont {Greiner}},
  \emph {et~al.},\ }\bibfield  {title} {\bibinfo {title} {Probing many-body
  dynamics on a 51-atom quantum simulator},\ }\href@noop {} {\bibfield
  {journal} {\bibinfo  {journal} {Nature}\ }\textbf {\bibinfo {volume} {551}},\
  \bibinfo {pages} {579} (\bibinfo {year} {2017})}\BibitemShut {NoStop}%
\bibitem [{\citenamefont {Keesling}\ \emph {et~al.}(2019)\citenamefont
  {Keesling}, \citenamefont {Omran}, \citenamefont {Levine}, \citenamefont
  {Bernien}, \citenamefont {Pichler}, \citenamefont {Choi}, \citenamefont
  {Samajdar}, \citenamefont {Schwartz}, \citenamefont {Silvi}, \citenamefont
  {Sachdev} \emph {et~al.}}]{keesling2019quantum}%
  \BibitemOpen
  \bibfield  {author} {\bibinfo {author} {\bibfnamefont {A.}~\bibnamefont
  {Keesling}}, \bibinfo {author} {\bibfnamefont {A.}~\bibnamefont {Omran}},
  \bibinfo {author} {\bibfnamefont {H.}~\bibnamefont {Levine}}, \bibinfo
  {author} {\bibfnamefont {H.}~\bibnamefont {Bernien}}, \bibinfo {author}
  {\bibfnamefont {H.}~\bibnamefont {Pichler}}, \bibinfo {author} {\bibfnamefont
  {S.}~\bibnamefont {Choi}}, \bibinfo {author} {\bibfnamefont {R.}~\bibnamefont
  {Samajdar}}, \bibinfo {author} {\bibfnamefont {S.}~\bibnamefont {Schwartz}},
  \bibinfo {author} {\bibfnamefont {P.}~\bibnamefont {Silvi}}, \bibinfo
  {author} {\bibfnamefont {S.}~\bibnamefont {Sachdev}}, \emph {et~al.},\
  }\bibfield  {title} {\bibinfo {title} {Quantum kibble--zurek mechanism and
  critical dynamics on a programmable rydberg simulator},\ }\href@noop {}
  {\bibfield  {journal} {\bibinfo  {journal} {Nature}\ }\textbf {\bibinfo
  {volume} {568}},\ \bibinfo {pages} {207} (\bibinfo {year}
  {2019})}\BibitemShut {NoStop}%
\bibitem [{\citenamefont {Scholl}\ \emph {et~al.}(2021)\citenamefont {Scholl},
  \citenamefont {Schuler}, \citenamefont {Williams}, \citenamefont
  {Eberharter}, \citenamefont {Barredo}, \citenamefont {Schymik}, \citenamefont
  {Lienhard}, \citenamefont {Henry}, \citenamefont {Lang}, \citenamefont
  {Lahaye} \emph {et~al.}}]{scholl2021quantum}%
  \BibitemOpen
  \bibfield  {author} {\bibinfo {author} {\bibfnamefont {P.}~\bibnamefont
  {Scholl}}, \bibinfo {author} {\bibfnamefont {M.}~\bibnamefont {Schuler}},
  \bibinfo {author} {\bibfnamefont {H.~J.}\ \bibnamefont {Williams}}, \bibinfo
  {author} {\bibfnamefont {A.~A.}\ \bibnamefont {Eberharter}}, \bibinfo
  {author} {\bibfnamefont {D.}~\bibnamefont {Barredo}}, \bibinfo {author}
  {\bibfnamefont {K.-N.}\ \bibnamefont {Schymik}}, \bibinfo {author}
  {\bibfnamefont {V.}~\bibnamefont {Lienhard}}, \bibinfo {author}
  {\bibfnamefont {L.-P.}\ \bibnamefont {Henry}}, \bibinfo {author}
  {\bibfnamefont {T.~C.}\ \bibnamefont {Lang}}, \bibinfo {author}
  {\bibfnamefont {T.}~\bibnamefont {Lahaye}}, \emph {et~al.},\ }\bibfield
  {title} {\bibinfo {title} {Quantum simulation of 2d antiferromagnets with
  hundreds of rydberg atoms},\ }\href@noop {} {\bibfield  {journal} {\bibinfo
  {journal} {Nature}\ }\textbf {\bibinfo {volume} {595}},\ \bibinfo {pages}
  {233} (\bibinfo {year} {2021})}\BibitemShut {NoStop}%
\bibitem [{\citenamefont {Ebadi}\ \emph {et~al.}(2021)\citenamefont {Ebadi},
  \citenamefont {Wang}, \citenamefont {Levine}, \citenamefont {Keesling},
  \citenamefont {Semeghini}, \citenamefont {Omran}, \citenamefont {Bluvstein},
  \citenamefont {Samajdar}, \citenamefont {Pichler}, \citenamefont {Ho} \emph
  {et~al.}}]{ebadi2021quantum}%
  \BibitemOpen
  \bibfield  {author} {\bibinfo {author} {\bibfnamefont {S.}~\bibnamefont
  {Ebadi}}, \bibinfo {author} {\bibfnamefont {T.~T.}\ \bibnamefont {Wang}},
  \bibinfo {author} {\bibfnamefont {H.}~\bibnamefont {Levine}}, \bibinfo
  {author} {\bibfnamefont {A.}~\bibnamefont {Keesling}}, \bibinfo {author}
  {\bibfnamefont {G.}~\bibnamefont {Semeghini}}, \bibinfo {author}
  {\bibfnamefont {A.}~\bibnamefont {Omran}}, \bibinfo {author} {\bibfnamefont
  {D.}~\bibnamefont {Bluvstein}}, \bibinfo {author} {\bibfnamefont
  {R.}~\bibnamefont {Samajdar}}, \bibinfo {author} {\bibfnamefont
  {H.}~\bibnamefont {Pichler}}, \bibinfo {author} {\bibfnamefont {W.~W.}\
  \bibnamefont {Ho}}, \emph {et~al.},\ }\bibfield  {title} {\bibinfo {title}
  {Quantum phases of matter on a 256-atom programmable quantum simulator},\
  }\href@noop {} {\bibfield  {journal} {\bibinfo  {journal} {Nature}\ }\textbf
  {\bibinfo {volume} {595}},\ \bibinfo {pages} {227} (\bibinfo {year}
  {2021})}\BibitemShut {NoStop}%
\bibitem [{\citenamefont {Bluvstein}\ \emph {et~al.}(2021)\citenamefont
  {Bluvstein}, \citenamefont {Omran}, \citenamefont {Levine}, \citenamefont
  {Keesling}, \citenamefont {Semeghini}, \citenamefont {Ebadi}, \citenamefont
  {Wang}, \citenamefont {Michailidis}, \citenamefont {Maskara}, \citenamefont
  {Ho} \emph {et~al.}}]{bluvstein2021controlling}%
  \BibitemOpen
  \bibfield  {author} {\bibinfo {author} {\bibfnamefont {D.}~\bibnamefont
  {Bluvstein}}, \bibinfo {author} {\bibfnamefont {A.}~\bibnamefont {Omran}},
  \bibinfo {author} {\bibfnamefont {H.}~\bibnamefont {Levine}}, \bibinfo
  {author} {\bibfnamefont {A.}~\bibnamefont {Keesling}}, \bibinfo {author}
  {\bibfnamefont {G.}~\bibnamefont {Semeghini}}, \bibinfo {author}
  {\bibfnamefont {S.}~\bibnamefont {Ebadi}}, \bibinfo {author} {\bibfnamefont
  {T.~T.}\ \bibnamefont {Wang}}, \bibinfo {author} {\bibfnamefont {A.~A.}\
  \bibnamefont {Michailidis}}, \bibinfo {author} {\bibfnamefont
  {N.}~\bibnamefont {Maskara}}, \bibinfo {author} {\bibfnamefont {W.~W.}\
  \bibnamefont {Ho}}, \emph {et~al.},\ }\bibfield  {title} {\bibinfo {title}
  {Controlling quantum many-body dynamics in driven rydberg atom arrays},\
  }\href@noop {} {\bibfield  {journal} {\bibinfo  {journal} {Science}\ }\textbf
  {\bibinfo {volume} {371}},\ \bibinfo {pages} {1355} (\bibinfo {year}
  {2021})}\BibitemShut {NoStop}%
\bibitem [{\citenamefont {Barredo}\ \emph {et~al.}(2015)\citenamefont
  {Barredo}, \citenamefont {Labuhn}, \citenamefont {Ravets}, \citenamefont
  {Lahaye}, \citenamefont {Browaeys},\ and\ \citenamefont
  {Adams}}]{barredo2015coherent}%
  \BibitemOpen
  \bibfield  {author} {\bibinfo {author} {\bibfnamefont {D.}~\bibnamefont
  {Barredo}}, \bibinfo {author} {\bibfnamefont {H.}~\bibnamefont {Labuhn}},
  \bibinfo {author} {\bibfnamefont {S.}~\bibnamefont {Ravets}}, \bibinfo
  {author} {\bibfnamefont {T.}~\bibnamefont {Lahaye}}, \bibinfo {author}
  {\bibfnamefont {A.}~\bibnamefont {Browaeys}},\ and\ \bibinfo {author}
  {\bibfnamefont {C.~S.}\ \bibnamefont {Adams}},\ }\bibfield  {title} {\bibinfo
  {title} {Coherent excitation transfer in a spin chain of three rydberg
  atoms},\ }\href@noop {} {\bibfield  {journal} {\bibinfo  {journal} {Physical
  review letters}\ }\textbf {\bibinfo {volume} {114}},\ \bibinfo {pages}
  {113002} (\bibinfo {year} {2015})}\BibitemShut {NoStop}%
\bibitem [{\citenamefont {Chen}\ \emph {et~al.}(2023)\citenamefont {Chen},
  \citenamefont {Bornet}, \citenamefont {Bintz}, \citenamefont {Emperauger},
  \citenamefont {Leclerc}, \citenamefont {Liu}, \citenamefont {Scholl},
  \citenamefont {Barredo}, \citenamefont {Hauschild}, \citenamefont
  {Chatterjee} \emph {et~al.}}]{chen2023continuous}%
  \BibitemOpen
  \bibfield  {author} {\bibinfo {author} {\bibfnamefont {C.}~\bibnamefont
  {Chen}}, \bibinfo {author} {\bibfnamefont {G.}~\bibnamefont {Bornet}},
  \bibinfo {author} {\bibfnamefont {M.}~\bibnamefont {Bintz}}, \bibinfo
  {author} {\bibfnamefont {G.}~\bibnamefont {Emperauger}}, \bibinfo {author}
  {\bibfnamefont {L.}~\bibnamefont {Leclerc}}, \bibinfo {author} {\bibfnamefont
  {V.~S.}\ \bibnamefont {Liu}}, \bibinfo {author} {\bibfnamefont
  {P.}~\bibnamefont {Scholl}}, \bibinfo {author} {\bibfnamefont
  {D.}~\bibnamefont {Barredo}}, \bibinfo {author} {\bibfnamefont
  {J.}~\bibnamefont {Hauschild}}, \bibinfo {author} {\bibfnamefont
  {S.}~\bibnamefont {Chatterjee}}, \emph {et~al.},\ }\bibfield  {title}
  {\bibinfo {title} {Continuous symmetry breaking in a two-dimensional rydberg
  array},\ }\href@noop {} {\bibfield  {journal} {\bibinfo  {journal} {Nature}\
  }\textbf {\bibinfo {volume} {616}},\ \bibinfo {pages} {691} (\bibinfo {year}
  {2023})}\BibitemShut {NoStop}%
\bibitem [{\citenamefont {de~L{\'e}s{\'e}leuc}\ \emph
  {et~al.}(2019)\citenamefont {de~L{\'e}s{\'e}leuc}, \citenamefont {Lienhard},
  \citenamefont {Scholl}, \citenamefont {Barredo}, \citenamefont {Weber},
  \citenamefont {Lang}, \citenamefont {B{\"u}chler}, \citenamefont {Lahaye},\
  and\ \citenamefont {Browaeys}}]{de2019observation}%
  \BibitemOpen
  \bibfield  {author} {\bibinfo {author} {\bibfnamefont {S.}~\bibnamefont
  {de~L{\'e}s{\'e}leuc}}, \bibinfo {author} {\bibfnamefont {V.}~\bibnamefont
  {Lienhard}}, \bibinfo {author} {\bibfnamefont {P.}~\bibnamefont {Scholl}},
  \bibinfo {author} {\bibfnamefont {D.}~\bibnamefont {Barredo}}, \bibinfo
  {author} {\bibfnamefont {S.}~\bibnamefont {Weber}}, \bibinfo {author}
  {\bibfnamefont {N.}~\bibnamefont {Lang}}, \bibinfo {author} {\bibfnamefont
  {H.~P.}\ \bibnamefont {B{\"u}chler}}, \bibinfo {author} {\bibfnamefont
  {T.}~\bibnamefont {Lahaye}},\ and\ \bibinfo {author} {\bibfnamefont
  {A.}~\bibnamefont {Browaeys}},\ }\bibfield  {title} {\bibinfo {title}
  {Observation of a symmetry-protected topological phase of interacting bosons
  with rydberg atoms},\ }\href@noop {} {\bibfield  {journal} {\bibinfo
  {journal} {Science}\ }\textbf {\bibinfo {volume} {365}},\ \bibinfo {pages}
  {775} (\bibinfo {year} {2019})}\BibitemShut {NoStop}%
\bibitem [{\citenamefont {Semeghini}\ \emph {et~al.}(2021)\citenamefont
  {Semeghini}, \citenamefont {Levine}, \citenamefont {Keesling}, \citenamefont
  {Ebadi}, \citenamefont {Wang}, \citenamefont {Bluvstein}, \citenamefont
  {Verresen}, \citenamefont {Pichler}, \citenamefont {Kalinowski},
  \citenamefont {Samajdar} \emph {et~al.}}]{semeghini2021probing}%
  \BibitemOpen
  \bibfield  {author} {\bibinfo {author} {\bibfnamefont {G.}~\bibnamefont
  {Semeghini}}, \bibinfo {author} {\bibfnamefont {H.}~\bibnamefont {Levine}},
  \bibinfo {author} {\bibfnamefont {A.}~\bibnamefont {Keesling}}, \bibinfo
  {author} {\bibfnamefont {S.}~\bibnamefont {Ebadi}}, \bibinfo {author}
  {\bibfnamefont {T.~T.}\ \bibnamefont {Wang}}, \bibinfo {author}
  {\bibfnamefont {D.}~\bibnamefont {Bluvstein}}, \bibinfo {author}
  {\bibfnamefont {R.}~\bibnamefont {Verresen}}, \bibinfo {author}
  {\bibfnamefont {H.}~\bibnamefont {Pichler}}, \bibinfo {author} {\bibfnamefont
  {M.}~\bibnamefont {Kalinowski}}, \bibinfo {author} {\bibfnamefont
  {R.}~\bibnamefont {Samajdar}}, \emph {et~al.},\ }\bibfield  {title} {\bibinfo
  {title} {Probing topological spin liquids on a programmable quantum
  simulator},\ }\href@noop {} {\bibfield  {journal} {\bibinfo  {journal}
  {Science}\ }\textbf {\bibinfo {volume} {374}},\ \bibinfo {pages} {1242}
  (\bibinfo {year} {2021})}\BibitemShut {NoStop}%
\bibitem [{\citenamefont {Yang}\ \emph {et~al.}(2022)\citenamefont {Yang},
  \citenamefont {Wang}, \citenamefont {Zhou},\ and\ \citenamefont
  {Liu}}]{PhysRevA.106.L021101}%
  \BibitemOpen
  \bibfield  {author} {\bibinfo {author} {\bibfnamefont {T.-H.}\ \bibnamefont
  {Yang}}, \bibinfo {author} {\bibfnamefont {B.-Z.}\ \bibnamefont {Wang}},
  \bibinfo {author} {\bibfnamefont {X.-C.}\ \bibnamefont {Zhou}},\ and\
  \bibinfo {author} {\bibfnamefont {X.-J.}\ \bibnamefont {Liu}},\ }\bibfield
  {title} {\bibinfo {title} {Quantum hall states for rydberg arrays with
  laser-assisted dipole-dipole interactions},\ }\href@noop {} {\bibfield
  {journal} {\bibinfo  {journal} {Phys. Rev. A}\ }\textbf {\bibinfo {volume}
  {106}},\ \bibinfo {pages} {L021101} (\bibinfo {year} {2022})}\BibitemShut
  {NoStop}%
\bibitem [{\citenamefont {Poon}\ \emph {et~al.}(2023)\citenamefont {Poon},
  \citenamefont {Zhou}, \citenamefont {Wang}, \citenamefont {Yang},\ and\
  \citenamefont {Liu}}]{poon2023fractional}%
  \BibitemOpen
  \bibfield  {author} {\bibinfo {author} {\bibfnamefont {T.-F.~J.}\
  \bibnamefont {Poon}}, \bibinfo {author} {\bibfnamefont {X.-C.}\ \bibnamefont
  {Zhou}}, \bibinfo {author} {\bibfnamefont {B.-Z.}\ \bibnamefont {Wang}},
  \bibinfo {author} {\bibfnamefont {T.-H.}\ \bibnamefont {Yang}},\ and\
  \bibinfo {author} {\bibfnamefont {X.-J.}\ \bibnamefont {Liu}},\ }\bibfield
  {title} {\bibinfo {title} {Fractional quantum anomalous hall phase for raman
  superarray of rydberg atoms},\ }\href@noop {} {\bibfield  {journal} {\bibinfo
   {journal} {arXiv preprint arXiv:2302.13104}\ } (\bibinfo {year}
  {2023})}\BibitemShut {NoStop}%
\bibitem [{\citenamefont {Zhou}\ \emph {et~al.}(2022)\citenamefont {Zhou},
  \citenamefont {Wang}, \citenamefont {Poon}, \citenamefont {Zhou},\ and\
  \citenamefont {Liu}}]{zhou2022exact}%
  \BibitemOpen
  \bibfield  {author} {\bibinfo {author} {\bibfnamefont {X.-C.}\ \bibnamefont
  {Zhou}}, \bibinfo {author} {\bibfnamefont {Y.}~\bibnamefont {Wang}}, \bibinfo
  {author} {\bibfnamefont {T.-F.~J.}\ \bibnamefont {Poon}}, \bibinfo {author}
  {\bibfnamefont {Q.}~\bibnamefont {Zhou}},\ and\ \bibinfo {author}
  {\bibfnamefont {X.-J.}\ \bibnamefont {Liu}},\ }\bibfield  {title} {\bibinfo
  {title} {Exact new mobility edges between critical and localized states},\
  }\href@noop {} {\bibfield  {journal} {\bibinfo  {journal} {arXiv preprint
  arXiv:2212.14285}\ } (\bibinfo {year} {2022})}\BibitemShut {NoStop}%
\bibitem [{\citenamefont {Aidelsburger}\ \emph {et~al.}(2013)\citenamefont
  {Aidelsburger}, \citenamefont {Atala}, \citenamefont {Lohse}, \citenamefont
  {Barreiro}, \citenamefont {Paredes},\ and\ \citenamefont
  {Bloch}}]{aidelsburger2013realization}%
  \BibitemOpen
  \bibfield  {author} {\bibinfo {author} {\bibfnamefont {M.}~\bibnamefont
  {Aidelsburger}}, \bibinfo {author} {\bibfnamefont {M.}~\bibnamefont {Atala}},
  \bibinfo {author} {\bibfnamefont {M.}~\bibnamefont {Lohse}}, \bibinfo
  {author} {\bibfnamefont {J.~T.}\ \bibnamefont {Barreiro}}, \bibinfo {author}
  {\bibfnamefont {B.}~\bibnamefont {Paredes}},\ and\ \bibinfo {author}
  {\bibfnamefont {I.}~\bibnamefont {Bloch}},\ }\bibfield  {title} {\bibinfo
  {title} {Realization of the hofstadter hamiltonian with ultracold atoms in
  optical lattices},\ }\href@noop {} {\bibfield  {journal} {\bibinfo  {journal}
  {Physical review letters}\ }\textbf {\bibinfo {volume} {111}},\ \bibinfo
  {pages} {185301} (\bibinfo {year} {2013})}\BibitemShut {NoStop}%
\bibitem [{\citenamefont {Miyake}\ \emph {et~al.}(2013)\citenamefont {Miyake},
  \citenamefont {Siviloglou}, \citenamefont {Kennedy}, \citenamefont {Burton},\
  and\ \citenamefont {Ketterle}}]{miyake2013realizing}%
  \BibitemOpen
  \bibfield  {author} {\bibinfo {author} {\bibfnamefont {H.}~\bibnamefont
  {Miyake}}, \bibinfo {author} {\bibfnamefont {G.~A.}\ \bibnamefont
  {Siviloglou}}, \bibinfo {author} {\bibfnamefont {C.~J.}\ \bibnamefont
  {Kennedy}}, \bibinfo {author} {\bibfnamefont {W.~C.}\ \bibnamefont
  {Burton}},\ and\ \bibinfo {author} {\bibfnamefont {W.}~\bibnamefont
  {Ketterle}},\ }\bibfield  {title} {\bibinfo {title} {Realizing the harper
  hamiltonian with laser-assisted tunneling in optical lattices},\ }\href@noop
  {} {\bibfield  {journal} {\bibinfo  {journal} {Physical review letters}\
  }\textbf {\bibinfo {volume} {111}},\ \bibinfo {pages} {185302} (\bibinfo
  {year} {2013})}\BibitemShut {NoStop}%
\bibitem [{\citenamefont {Liu}\ \emph {et~al.}(2014)\citenamefont {Liu},
  \citenamefont {Law},\ and\ \citenamefont {Ng}}]{liu2014realization}%
  \BibitemOpen
  \bibfield  {author} {\bibinfo {author} {\bibfnamefont {X.-J.}\ \bibnamefont
  {Liu}}, \bibinfo {author} {\bibfnamefont {K.~T.}\ \bibnamefont {Law}},\ and\
  \bibinfo {author} {\bibfnamefont {T.~K.}\ \bibnamefont {Ng}},\ }\bibfield
  {title} {\bibinfo {title} {Realization of 2d spin-orbit interaction and
  exotic topological orders in cold atoms},\ }\href@noop {} {\bibfield
  {journal} {\bibinfo  {journal} {Physical Review Letters}\ }\textbf {\bibinfo
  {volume} {112}},\ \bibinfo {pages} {086401} (\bibinfo {year}
  {2014})}\BibitemShut {NoStop}%
\bibitem [{\citenamefont {Aidelsburger}\ \emph {et~al.}(2015)\citenamefont
  {Aidelsburger}, \citenamefont {Lohse}, \citenamefont {Schweizer},
  \citenamefont {Atala}, \citenamefont {Barreiro}, \citenamefont
  {Nascimb{\`e}ne}, \citenamefont {Cooper}, \citenamefont {Bloch},\ and\
  \citenamefont {Goldman}}]{aidelsburger2015measuring}%
  \BibitemOpen
  \bibfield  {author} {\bibinfo {author} {\bibfnamefont {M.}~\bibnamefont
  {Aidelsburger}}, \bibinfo {author} {\bibfnamefont {M.}~\bibnamefont {Lohse}},
  \bibinfo {author} {\bibfnamefont {C.}~\bibnamefont {Schweizer}}, \bibinfo
  {author} {\bibfnamefont {M.}~\bibnamefont {Atala}}, \bibinfo {author}
  {\bibfnamefont {J.~T.}\ \bibnamefont {Barreiro}}, \bibinfo {author}
  {\bibfnamefont {S.}~\bibnamefont {Nascimb{\`e}ne}}, \bibinfo {author}
  {\bibfnamefont {N.}~\bibnamefont {Cooper}}, \bibinfo {author} {\bibfnamefont
  {I.}~\bibnamefont {Bloch}},\ and\ \bibinfo {author} {\bibfnamefont
  {N.}~\bibnamefont {Goldman}},\ }\bibfield  {title} {\bibinfo {title}
  {Measuring the chern number of hofstadter bands with ultracold bosonic
  atoms},\ }\href@noop {} {\bibfield  {journal} {\bibinfo  {journal} {Nature
  Physics}\ }\textbf {\bibinfo {volume} {11}},\ \bibinfo {pages} {162}
  (\bibinfo {year} {2015})}\BibitemShut {NoStop}%
\bibitem [{\citenamefont {Wu}\ \emph {et~al.}(2016)\citenamefont {Wu},
  \citenamefont {Zhang}, \citenamefont {Sun}, \citenamefont {Xu}, \citenamefont
  {Wang}, \citenamefont {Ji}, \citenamefont {Deng}, \citenamefont {Chen},
  \citenamefont {Liu},\ and\ \citenamefont {Pan}}]{wu2016realization}%
  \BibitemOpen
  \bibfield  {author} {\bibinfo {author} {\bibfnamefont {Z.}~\bibnamefont
  {Wu}}, \bibinfo {author} {\bibfnamefont {L.}~\bibnamefont {Zhang}}, \bibinfo
  {author} {\bibfnamefont {W.}~\bibnamefont {Sun}}, \bibinfo {author}
  {\bibfnamefont {X.-T.}\ \bibnamefont {Xu}}, \bibinfo {author} {\bibfnamefont
  {B.-Z.}\ \bibnamefont {Wang}}, \bibinfo {author} {\bibfnamefont {S.-C.}\
  \bibnamefont {Ji}}, \bibinfo {author} {\bibfnamefont {Y.}~\bibnamefont
  {Deng}}, \bibinfo {author} {\bibfnamefont {S.}~\bibnamefont {Chen}}, \bibinfo
  {author} {\bibfnamefont {X.-J.}\ \bibnamefont {Liu}},\ and\ \bibinfo {author}
  {\bibfnamefont {J.-W.}\ \bibnamefont {Pan}},\ }\bibfield  {title} {\bibinfo
  {title} {Realization of two-dimensional spin-orbit coupling for bose-einstein
  condensates},\ }\href@noop {} {\bibfield  {journal} {\bibinfo  {journal}
  {Science}\ }\textbf {\bibinfo {volume} {354}},\ \bibinfo {pages} {83}
  (\bibinfo {year} {2016})}\BibitemShut {NoStop}%
\bibitem [{\citenamefont {Liu}\ \emph {et~al.}(2016)\citenamefont {Liu},
  \citenamefont {Liu}, \citenamefont {Law}, \citenamefont {Liu},\ and\
  \citenamefont {Ng}}]{liu2016chiral}%
  \BibitemOpen
  \bibfield  {author} {\bibinfo {author} {\bibfnamefont {X.-J.}\ \bibnamefont
  {Liu}}, \bibinfo {author} {\bibfnamefont {Z.-X.}\ \bibnamefont {Liu}},
  \bibinfo {author} {\bibfnamefont {K.~T.}\ \bibnamefont {Law}}, \bibinfo
  {author} {\bibfnamefont {W.~V.}\ \bibnamefont {Liu}},\ and\ \bibinfo {author}
  {\bibfnamefont {T.~K.}\ \bibnamefont {Ng}},\ }\bibfield  {title} {\bibinfo
  {title} {Chiral topological orders in an optical raman lattice},\ }\href@noop
  {} {\bibfield  {journal} {\bibinfo  {journal} {New Journal of Physics}\
  }\textbf {\bibinfo {volume} {18}},\ \bibinfo {pages} {035004} (\bibinfo
  {year} {2016})}\BibitemShut {NoStop}%
\bibitem [{\citenamefont {Song}\ \emph {et~al.}(2018)\citenamefont {Song},
  \citenamefont {Zhang}, \citenamefont {He}, \citenamefont {Poon},
  \citenamefont {Hajiyev}, \citenamefont {Zhang}, \citenamefont {Liu},\ and\
  \citenamefont {Jo}}]{song2018observation}%
  \BibitemOpen
  \bibfield  {author} {\bibinfo {author} {\bibfnamefont {B.}~\bibnamefont
  {Song}}, \bibinfo {author} {\bibfnamefont {L.}~\bibnamefont {Zhang}},
  \bibinfo {author} {\bibfnamefont {C.}~\bibnamefont {He}}, \bibinfo {author}
  {\bibfnamefont {T.~F.~J.}\ \bibnamefont {Poon}}, \bibinfo {author}
  {\bibfnamefont {E.}~\bibnamefont {Hajiyev}}, \bibinfo {author} {\bibfnamefont
  {S.}~\bibnamefont {Zhang}}, \bibinfo {author} {\bibfnamefont {X.-J.}\
  \bibnamefont {Liu}},\ and\ \bibinfo {author} {\bibfnamefont {G.-B.}\
  \bibnamefont {Jo}},\ }\bibfield  {title} {\bibinfo {title} {Observation of
  symmetry-protected topological band with ultracold fermions},\ }\href@noop {}
  {\bibfield  {journal} {\bibinfo  {journal} {Science advances}\ }\textbf
  {\bibinfo {volume} {4}},\ \bibinfo {pages} {eaao4748} (\bibinfo {year}
  {2018})}\BibitemShut {NoStop}%
\bibitem [{\citenamefont {Lu}\ \emph {et~al.}(2020)\citenamefont {Lu},
  \citenamefont {Wang},\ and\ \citenamefont {Liu}}]{lu2020ideal}%
  \BibitemOpen
  \bibfield  {author} {\bibinfo {author} {\bibfnamefont {Y.-H.}\ \bibnamefont
  {Lu}}, \bibinfo {author} {\bibfnamefont {B.-Z.}\ \bibnamefont {Wang}},\ and\
  \bibinfo {author} {\bibfnamefont {X.-J.}\ \bibnamefont {Liu}},\ }\bibfield
  {title} {\bibinfo {title} {Ideal weyl semimetal with 3d spin-orbit coupled
  ultracold quantum gas},\ }\href@noop {} {\bibfield  {journal} {\bibinfo
  {journal} {Science Bulletin}\ }\textbf {\bibinfo {volume} {65}},\ \bibinfo
  {pages} {2080} (\bibinfo {year} {2020})}\BibitemShut {NoStop}%
\bibitem [{\citenamefont {Wang}\ \emph {et~al.}(2021)\citenamefont {Wang},
  \citenamefont {Cheng}, \citenamefont {Wang}, \citenamefont {Zhang},
  \citenamefont {Lu}, \citenamefont {Yi}, \citenamefont {Niu}, \citenamefont
  {Deng}, \citenamefont {Liu}, \citenamefont {Chen} \emph
  {et~al.}}]{wang2021realization}%
  \BibitemOpen
  \bibfield  {author} {\bibinfo {author} {\bibfnamefont {Z.-Y.}\ \bibnamefont
  {Wang}}, \bibinfo {author} {\bibfnamefont {X.-C.}\ \bibnamefont {Cheng}},
  \bibinfo {author} {\bibfnamefont {B.-Z.}\ \bibnamefont {Wang}}, \bibinfo
  {author} {\bibfnamefont {J.-Y.}\ \bibnamefont {Zhang}}, \bibinfo {author}
  {\bibfnamefont {Y.-H.}\ \bibnamefont {Lu}}, \bibinfo {author} {\bibfnamefont
  {C.-R.}\ \bibnamefont {Yi}}, \bibinfo {author} {\bibfnamefont
  {S.}~\bibnamefont {Niu}}, \bibinfo {author} {\bibfnamefont {Y.}~\bibnamefont
  {Deng}}, \bibinfo {author} {\bibfnamefont {X.-J.}\ \bibnamefont {Liu}},
  \bibinfo {author} {\bibfnamefont {S.}~\bibnamefont {Chen}}, \emph {et~al.},\
  }\bibfield  {title} {\bibinfo {title} {Realization of an ideal weyl semimetal
  band in a quantum gas with 3d spin-orbit coupling},\ }\href@noop {}
  {\bibfield  {journal} {\bibinfo  {journal} {Science}\ }\textbf {\bibinfo
  {volume} {372}},\ \bibinfo {pages} {271} (\bibinfo {year}
  {2021})}\BibitemShut {NoStop}%
\bibitem [{Note1()}]{Note1}%
  \BibitemOpen
  \bibinfo {note} {$J_{+-}^{\gamma }$ ($J_{--}^{\gamma }$) denote hopping
  (pairing) term across $\gamma $ bond.}\BibitemShut {Stop}%
\bibitem [{Note2()}]{Note2}%
  \BibitemOpen
  \bibinfo {note} {See more details in Supplementary material about the laser
  assisted dipole-dipole interaction, numerical estimate of parameters, and the
  detection schemes.}\BibitemShut {Stop}%
\bibitem [{\citenamefont {Zhu}\ \emph {et~al.}(2018)\citenamefont {Zhu},
  \citenamefont {Kimchi}, \citenamefont {Sheng},\ and\ \citenamefont
  {Fu}}]{zhu2018robust}%
  \BibitemOpen
  \bibfield  {author} {\bibinfo {author} {\bibfnamefont {Z.}~\bibnamefont
  {Zhu}}, \bibinfo {author} {\bibfnamefont {I.}~\bibnamefont {Kimchi}},
  \bibinfo {author} {\bibfnamefont {D.}~\bibnamefont {Sheng}},\ and\ \bibinfo
  {author} {\bibfnamefont {L.}~\bibnamefont {Fu}},\ }\bibfield  {title}
  {\bibinfo {title} {Robust non-abelian spin liquid and a possible intermediate
  phase in the antiferromagnetic kitaev model with magnetic field},\
  }\href@noop {} {\bibfield  {journal} {\bibinfo  {journal} {Physical Review
  B}\ }\textbf {\bibinfo {volume} {97}},\ \bibinfo {pages} {241110} (\bibinfo
  {year} {2018})}\BibitemShut {NoStop}%
\bibitem [{\citenamefont {Knolle}\ \emph {et~al.}(2014)\citenamefont {Knolle},
  \citenamefont {Chern}, \citenamefont {Kovrizhin}, \citenamefont {Moessner},\
  and\ \citenamefont {Perkins}}]{knolle2014raman}%
  \BibitemOpen
  \bibfield  {author} {\bibinfo {author} {\bibfnamefont {J.}~\bibnamefont
  {Knolle}}, \bibinfo {author} {\bibfnamefont {G.-W.}\ \bibnamefont {Chern}},
  \bibinfo {author} {\bibfnamefont {D.}~\bibnamefont {Kovrizhin}}, \bibinfo
  {author} {\bibfnamefont {R.}~\bibnamefont {Moessner}},\ and\ \bibinfo
  {author} {\bibfnamefont {N.}~\bibnamefont {Perkins}},\ }\bibfield  {title}
  {\bibinfo {title} {Raman scattering signatures of kitaev spin liquids in a 2
  iro 3 iridates with a= na or li},\ }\href@noop {} {\bibfield  {journal}
  {\bibinfo  {journal} {Physical review letters}\ }\textbf {\bibinfo {volume}
  {113}},\ \bibinfo {pages} {187201} (\bibinfo {year} {2014})}\BibitemShut
  {NoStop}%
\bibitem [{\citenamefont {Liu}\ \emph {et~al.}(2010)\citenamefont {Liu},
  \citenamefont {Liu}, \citenamefont {Wu},\ and\ \citenamefont
  {Sinova}}]{liu2010quantum}%
  \BibitemOpen
  \bibfield  {author} {\bibinfo {author} {\bibfnamefont {X.-J.}\ \bibnamefont
  {Liu}}, \bibinfo {author} {\bibfnamefont {X.}~\bibnamefont {Liu}}, \bibinfo
  {author} {\bibfnamefont {C.}~\bibnamefont {Wu}},\ and\ \bibinfo {author}
  {\bibfnamefont {J.}~\bibnamefont {Sinova}},\ }\bibfield  {title} {\bibinfo
  {title} {Quantum anomalous hall effect with cold atoms trapped in a square
  lattice},\ }\href@noop {} {\bibfield  {journal} {\bibinfo  {journal}
  {Physical Review A}\ }\textbf {\bibinfo {volume} {81}},\ \bibinfo {pages}
  {033622} (\bibinfo {year} {2010})}\BibitemShut {NoStop}%
\bibitem [{\citenamefont {Stanescu}\ \emph {et~al.}(2010)\citenamefont
  {Stanescu}, \citenamefont {Galitski},\ and\ \citenamefont
  {Sarma}}]{stanescu2010topological}%
  \BibitemOpen
  \bibfield  {author} {\bibinfo {author} {\bibfnamefont {T.~D.}\ \bibnamefont
  {Stanescu}}, \bibinfo {author} {\bibfnamefont {V.}~\bibnamefont {Galitski}},\
  and\ \bibinfo {author} {\bibfnamefont {S.~D.}\ \bibnamefont {Sarma}},\
  }\bibfield  {title} {\bibinfo {title} {Topological states in two-dimensional
  optical lattices},\ }\href@noop {} {\bibfield  {journal} {\bibinfo  {journal}
  {Physical Review A}\ }\textbf {\bibinfo {volume} {82}},\ \bibinfo {pages}
  {013608} (\bibinfo {year} {2010})}\BibitemShut {NoStop}%
\bibitem [{\citenamefont {Goldman}\ \emph {et~al.}(2012)\citenamefont
  {Goldman}, \citenamefont {Beugnon},\ and\ \citenamefont
  {Gerbier}}]{goldman2012detecting}%
  \BibitemOpen
  \bibfield  {author} {\bibinfo {author} {\bibfnamefont {N.}~\bibnamefont
  {Goldman}}, \bibinfo {author} {\bibfnamefont {J.}~\bibnamefont {Beugnon}},\
  and\ \bibinfo {author} {\bibfnamefont {F.}~\bibnamefont {Gerbier}},\
  }\bibfield  {title} {\bibinfo {title} {Detecting chiral edge states in the
  hofstadter optical lattice},\ }\href@noop {} {\bibfield  {journal} {\bibinfo
  {journal} {Physical review letters}\ }\textbf {\bibinfo {volume} {108}},\
  \bibinfo {pages} {255303} (\bibinfo {year} {2012})}\BibitemShut {NoStop}%
\bibitem [{\citenamefont {Buchhold}\ \emph {et~al.}(2012)\citenamefont
  {Buchhold}, \citenamefont {Cocks},\ and\ \citenamefont
  {Hofstetter}}]{buchhold2012effects}%
  \BibitemOpen
  \bibfield  {author} {\bibinfo {author} {\bibfnamefont {M.}~\bibnamefont
  {Buchhold}}, \bibinfo {author} {\bibfnamefont {D.}~\bibnamefont {Cocks}},\
  and\ \bibinfo {author} {\bibfnamefont {W.}~\bibnamefont {Hofstetter}},\
  }\bibfield  {title} {\bibinfo {title} {Effects of smooth boundaries on
  topological edge modes in optical lattices},\ }\href@noop {} {\bibfield
  {journal} {\bibinfo  {journal} {Physical Review A}\ }\textbf {\bibinfo
  {volume} {85}},\ \bibinfo {pages} {063614} (\bibinfo {year}
  {2012})}\BibitemShut {NoStop}%
\bibitem [{\citenamefont {Walker}\ and\ \citenamefont
  {Saffman}(2008)}]{walker2008consequences}%
  \BibitemOpen
  \bibfield  {author} {\bibinfo {author} {\bibfnamefont {T.~G.}\ \bibnamefont
  {Walker}}\ and\ \bibinfo {author} {\bibfnamefont {M.}~\bibnamefont
  {Saffman}},\ }\bibfield  {title} {\bibinfo {title} {Consequences of zeeman
  degeneracy for the van der waals blockade between rydberg atoms},\
  }\href@noop {} {\bibfield  {journal} {\bibinfo  {journal} {Physical Review
  A}\ }\textbf {\bibinfo {volume} {77}},\ \bibinfo {pages} {032723} (\bibinfo
  {year} {2008})}\BibitemShut {NoStop}%
\bibitem [{\citenamefont {Lieb}(1994)}]{lieb1994flux}%
  \BibitemOpen
  \bibfield  {author} {\bibinfo {author} {\bibfnamefont {E.~H.}\ \bibnamefont
  {Lieb}},\ }\bibfield  {title} {\bibinfo {title} {Flux phase of the
  half-filled band},\ }\href@noop {} {\bibfield  {journal} {\bibinfo  {journal}
  {Physical review letters}\ }\textbf {\bibinfo {volume} {73}},\ \bibinfo
  {pages} {2158} (\bibinfo {year} {1994})}\BibitemShut {NoStop}%
\bibitem [{\citenamefont {Haldane}(1988)}]{haldane1988model}%
  \BibitemOpen
  \bibfield  {author} {\bibinfo {author} {\bibfnamefont {F.~D.~M.}\
  \bibnamefont {Haldane}},\ }\bibfield  {title} {\bibinfo {title} {Model for a
  quantum hall effect without landau levels: Condensed-matter realization of
  the" parity anomaly"},\ }\href@noop {} {\bibfield  {journal} {\bibinfo
  {journal} {Physical review letters}\ }\textbf {\bibinfo {volume} {61}},\
  \bibinfo {pages} {2015} (\bibinfo {year} {1988})}\BibitemShut {NoStop}%
\end{thebibliography}%

%%%%%%%%%%%%%%%%%%%%%%%%%%%%%%%%%%%%%% %%   Supplementary Information %%%%%%%%%%%%%%%%%%%%%%%%%%%%%%%%%%%%%%
\renewcommand{\thesection}{S-\arabic{section}}
\setcounter{section}{0}  %  this will re-count section from 1
\renewcommand{\theequation}{S\arabic{equation}}
\setcounter{equation}{0}  %  this will re-count eq from 1
\renewcommand{\thefigure}{S\arabic{figure}}
\setcounter{figure}{0}  %  this will re-count eq from 1
\renewcommand{\thetable}{S\Roman{table}}
\setcounter{table}{0}  %  this will re-count eq from 1
\onecolumngrid \flushbottom %\onecolumn

\newpage
\begin{center}
\large \textbf{\large Supplementary Material: Realization and detection of Kitaev quantum spin liquid with Rydberg atoms}
\end{center}

\section{Derivation of laser-assisted interaction}\label{laddi}
In this section, we derive the effective hopping exchange term $J_{+-}$ and pairing exchange term $J_{--}$ using time-dependent perturbation theory.
The initial Hamiltonian is expressed as $H = H_{\mathrm{dipole}} + H_{\Delta} + V_{\mathrm{exchange}}(\mathbf{r},t) + V_{\mathrm{pairing}}(\mathbf{r},t)$, where $H_{\mathrm{dipole}}$ and $H_{\Delta}$ denote the bare dipole-dipole interaction and the on-site detuning, respectively. Two Raman potentials  $V_{\mathrm{exchange}}(\mathbf{r},t)$ and $V_{\mathrm{pairing}}(\mathbf{r},t)$ are described as
\begin{equation}
\left\{
\begin{aligned} &V_{\mathrm{exchange}}(\mathbf{r},t) = \sum_j \left(\Omega_1(\mathbf{r}_j) e^{-i \omega_1 t}+\Omega_e(\mathbf{r}_j) e^{-i \omega_e t}\right) \left| \downarrow  \right\rangle_{j_x, j_y}\left\langle 6P \right|_{j_x, j_y}+\text { h.c. }  \\
&V_{\mathrm{pairing}}(\mathbf{r},t) = \sum_j \left(\Omega_2(\mathbf{r}_j)  e^{-i \omega_2 t}+\Omega_p(\mathbf{r}_j) e^{-i \omega_p t}\right) \left| \uparrow \right\rangle_{j_x, j_y} \left\langle 5D \right|_{j_x, j_y} + \text{ h.c. }
\end{aligned}
\right.
\end{equation}
We focus on two neighboring sites $i$ and $j$ to elucidate the exchange and pairing processes. We label $|a\rangle=\left|\uparrow\right\rangle_i \left|\downarrow \right\rangle_j$ and  $|b\rangle=\left|\downarrow \right\rangle_i\left|\uparrow\right\rangle_j$ for convenience.
The exchange transitions between $|a\rangle$ and $|b\rangle$ are initially suppressed by the large on-site detunings $\Delta_e=(E_{\uparrow}-E_{\downarrow})_j-(E_{\uparrow}-E_{\downarrow})_i$, but are further recovered via the application of the Raman coupling potential $V_{\mathrm{exchange}}$.
The pairing process between $|c\rangle=\left|\uparrow\right\rangle_i\left|\uparrow\right\rangle_j$ and $|d\rangle=\left|\downarrow\right\rangle_i \left|\downarrow\right\rangle_j$ is naturally forbidden, it is subsequently facilitated by the Raman potential $V_{\mathrm{pairing}}$.

As illustrated in Fig.~\ref{Fig1}(c) of the main text, the Raman potential $V_{\mathrm{exchange}}$ couples the spin-down state $\left| \downarrow \right\rangle_j$ to the intermediate state $| 6P \rangle_j$ with $F=3/2$. Consequently, the exchange process can be described within a three-dimensional subspace spanned by $\left| a \right\rangle$, $\left| b \right\rangle$, and $\left| i \right\rangle_e = \left|\downarrow \right\rangle_i | 6P \rangle_j$. The Raman potential on site $i$ need not be considered here, since in our setup, Raman coupling has frequency matching the detuning only on one end of a bond. Then the effective Hamiltonian under the basis ($\left| a \right\rangle$, $\left| i \right\rangle_e$, $\left| b \right\rangle$) can be expressed as
\begin{equation}
H_{\mathrm{eff,exchange}}=\left(\begin{array}{ccc}
    E_{a} & \Omega_1(\mathbf{r}_j) e^{-i \omega_1 t}+\Omega_e (\mathbf{r}_j) e^{-i \omega_e t} & J_{d1}\\
   \Omega^*_1(\mathbf{r}_j) e^{i \omega_1 t}+\Omega^*_e(\mathbf{r}_j) e^{i \omega_e t} & E_{i_e} & \\
    J_{d1} &  & E_{b}
\end{array}\right). \label{exchange-perturbation}
\end{equation}
The diagonal terms correspond to the energies of the three states, satisfying: $E_b-E_a=\Delta_e$ and $E_{i_e}-E_a=\delta_1-\omega_1$. Using the diagonal matrix $U=diag(1, e^{-i(E_{i_e}-E_a)t}, e^{-i(E_b-E_a)t})$ to transform the Hamiltonian to the rotating frame we have
\begin{equation}
H_{\mathrm{eff,exchange}}^{(\mathrm{rot})}=\left(\begin{array}{ccc}
    0 & \Omega_{1}(\mathbf{r}_j)e^{-i\delta_1 t}+\Omega_e(\mathbf{r}_j)e^{-i(\delta_1-\omega_1
    +\omega_e)t} & J_{d1}e^{-i\Delta_e t}\\
    \Omega_{1}^{\ast}(\mathbf{r}_j)e^{i\delta_1 t}+\Omega_e^{\ast}(\mathbf{r}_j)e^{i(\delta_1-\omega_1
    +\omega_e)t} & 0 & 0\\J_{d1}e^{i\Delta_e t} & 0 & 0
\end{array}\right).
\end{equation}
From the time-dependent perturbation  theory of the Dyson series, the unitary evolution operator should be $U(t)=\mathscr{T} \exp \left[ -\frac{i}{\hbar} \int_0^t \mathrm{d} t_1 H(t_1) \right]$. Here we consider the possible perturbation terms that contribute to the transition from the state $\left| a \right\rangle$ to $\left| b \right\rangle$.
The first-order process, referred to as the bare dipole-dipole interaction, remains off-resonance. The second-order process is absent for this transition. In the third-order process, however, a resonant transition emerges through the following perturbation process
\begin{align}
U^{(3)} (t) =& - \left(\frac{i}{\hbar}\right)^3 \int_0^t \mathrm{d}t_1 \int_0^{t_1} \mathrm{d}t_2 \int_0^{t_3} \mathrm{d}t_3 \  H_{31}(t_1) H_{12}(t_2) H_{21}(t_3) \nonumber \\
=& - \left(\frac{i}{\hbar}\right)^3 \frac{\Omega_1(\mathbf{r}_j)\Omega^\ast_e(\mathbf{r}_j) J_{d1}}{\delta_1 \Delta_e (\Delta_e+\omega_1-\omega_e)} \left(e^{-i(\Delta_e+\omega_1-\omega_e)t} - 1 \right) + \cdots
\end{align}
In the last line, we have neglected non-resonant terms and retained only those that are resonant. The outcome reveals a Rabi transition characterized by a detuning of  $\Delta_e + \omega_1-\omega_e$. The Rabi transition is resonant when the condition $\Delta_e = \omega_e-\omega_1$ is satisfied, resulting in a resonant amplitude of
\begin{align}\label{sup_exchange_eq}
J_{+-} = \frac{\Omega_1^\ast(\mathbf{r}_j) \Omega_e(\mathbf{r}_j)}{\delta_1\Delta_e} J_{d1}+O((J_{d1},U_e)^4) = \frac{U_e(\mathbf{r}_j)}{\Delta_e} J_{d1}+O((J_{d1},U_e)^4).
\end{align}
At the end of Eq. (\ref{sup_exchange_eq}), we append the next order corrections. The fourth-order term makes no contribution to this transition, but it is worth noting that some fifth-order perturbations could influence this transition:
\begin{align}
U^{(5)} (t) =& - \left(\frac{i}{\hbar}\right)^5 \int_0^t \mathrm{d}t_1 \cdots \int_0^{t_4} \mathrm{d}t_5 \left[ H_{31}(t_1)H_{13}(t_2)H_{31}(t_3) H_{12}(t_4) H_{21}(t_5) +  H_{31}(t_1)H_{12}(t_2)H_{21}(t_3) H_{12}(t_4) H_{21}(t_5) \right] \nonumber \\
=& - \left(\frac{i}{\hbar}\right)^5 \frac{\Omega_1(\mathbf{r}_j)\Omega^\ast_2(\mathbf{r}_j) J_{d1}^3}{ (\Delta_e+\omega_1-\omega_e) \delta_1 \Delta_e^3} \left(e^{-i(\Delta_e+\omega_1-\omega_e)t} - 1 \right) + \cdots
\end{align}
This calculation reveals that higher-order corrections can arise from the multiple dipole-dipole interactions and multiple two-photon compensations.
The presence of multiple dipole-dipole interactions introduces a correction factor $\eta_1 = J_{d1}^2/\Delta_e^2$ that multiplies $J_{+-}$, while the occurrence of multiple two-photon compensations results in a correction factor $\eta_2={U_e^2}/{\Delta_e^2}$.
These correction factors  $\eta_1$ and $\eta_2$ only affect the amplitude of $J_{+-}$ without influencing its phase. This is  because the dipole-dipole interaction  inherently lacks the phase and any additional phase introduced by the two-photon process effectively cancels out.
If $\eta_1$ or $\eta_2$ has a finite value, higher-order corrections may significantly alter the leading order.
Fortunately, in our experimental setup, we estimate $J_{d1}$ and $U_e$ to be less than $0.2\Delta_e$, yielding  $\eta_{1/2} < 1/25$.
As a result, the subleading order, which is one to two orders of magnitude smaller than the leading order, can be safely neglected.
% Similarly, when Raman potential on $i$ assists resonant transition, the resulting coupling is given by $J_{+-}=-\frac{U_e(\mathbf{r}_i)}{\Delta_e}J_{d1}$.

In a similar manner, the pairing process is facilitated by the Raman potential $V_{\mathrm{pairing}}$, which couples the spin-up state $\left| \uparrow \right\rangle$ to the intermediate state $\left| 5D \right\rangle$ with $F=5/2$. Similar to the previous case, we only need to consider the site $j$ here. Then the effective
Hamiltonian under the basis ($\left| c \right\rangle$, $\left| i \right\rangle_{p,1}=\left|\uparrow\right\rangle_i |5D \rangle_j$,$\left| i \right\rangle_{p,2}=\left|\uparrow\right\rangle_i\left|F\right\rangle_j$, $\left| d \right\rangle$) can be written as
\begin{equation}
H_{\mathrm{eff,pairing}}=\left(\begin{array}{cccc}
    E_c & \Omega_2(\mathbf{r}_j) e^{-i \omega_2 t}+\Omega_p (\mathbf{r}_j) e^{-i \omega_p t} & 0 & 0\\
   \Omega^*_2(\mathbf{r}_j) e^{i \omega_2 t}+\Omega^*_p(\mathbf{r}_j) e^{i \omega_p t} & E_{i_{p,1}} & \Omega^*_2(\mathbf{r}_j) e^{i \omega_2 t}+\Omega^*_p(\mathbf{r}_j) e^{i \omega_p t} & 0\\
    0 & \Omega_2(\mathbf{r}_j) e^{-i \omega_2 t}+\Omega_p (\mathbf{r}_j) e^{-i \omega_p t} & E_{i_{p,2}} & J_{d2}\\
    0 & 0 & J_{d2} & E_d
\end{array}\right). \label{pairing-perturbation}
\end{equation}
To assist resonant transition, we choose $\omega_2-\omega_p=E_c-E_d$. Through a transformation to the rotating frame and employing the same perturbation method, we derive an resonant transition amplitude from the third-order perturbation as
\begin{align}
U^{(3)} (t) =& - \left(\frac{i}{\hbar}\right)^3 \int_0^t \mathrm{d}t_1 \int_0^{t_1} \mathrm{d}t_2 \int_0^{t_3} \mathrm{d}t_3 \  H_{43}(t_1) H_{32}(t_2) H_{21}(t_3) \nonumber \\
=& - \left(\frac{i}{\hbar}\right)^3 \frac{\Omega_2(\mathbf{r}_j)\Omega^\ast_p(\mathbf{r}_j) J_{d1}}{\delta_2 \Delta_p (E_c-E_d-\omega_2+\omega_p)} \left(e^{-i(E_c-E_d-\omega_2+\omega_p)t} - 1 \right) + \cdots
\end{align}
keeping only the resonant part. This amplitude correspondes to Rabi transition with detuning, and
\begin{align}\label{sup_pairing_eq}
J_{--} = \frac{\Omega_2^\ast(\mathbf{r}_j) \Omega_p(\mathbf{r}_j)}{\delta_2\Delta_p} J_{d2}+O((J_{d2},U_p)^4) = \frac{U_p(\mathbf{r}_j)}{\Delta_p} J_{d2}+O((J_{d2},U_p)^4).
\end{align}
We also append the next order corrections in (\ref{sup_pairing_eq}). Similarly to exchange process, we can identify higher-order contributions introducing two correction factors $\eta_1$ and $\eta_2$. However, based on the similar estimations of $J_{d2},U_p<0.2\Delta_p$, resulting in $\eta_{1/2} < 1/25$, the subleading order terms are small enough to be neglected.

\section{Experimental parameters}
\begin{figure}[H]
\centering
\includegraphics[width=0.8\columnwidth]{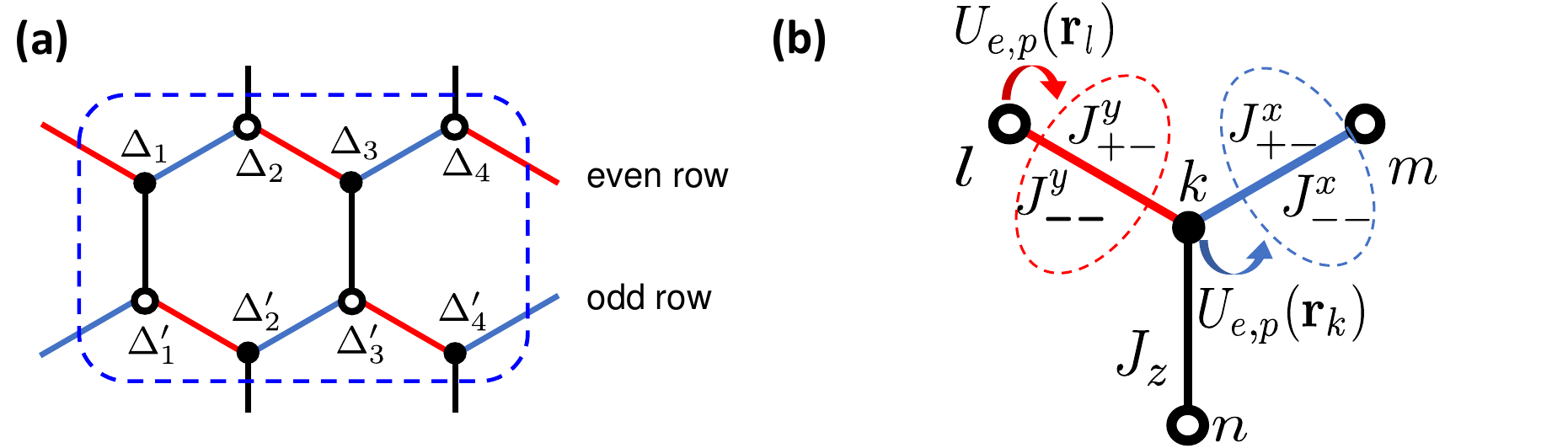}
\par
\centering{}\caption{(a) The 8-site periodic configuration of on-site energy shift $\Delta$ and Raman potential $U_{e,p}$  (on-site Raman potentials have the same site labels as detunings). (b) The coupling at each bond is only determined by the Raman potential on the right side. For instance, the coupling $J_{+-,--}$ at the $\langle kl\rangle$ bond is assisted by $U_{e(p)}(\mathbf{r}_l)$, while at the $\langle km\rangle$ bond it is assisted by $U_{e(p)}(\mathbf{r}_k)$.  }\label{periodic configuration}
\end{figure}

In this section we provide an estimation of experimental parameters based on the $^{87}$Rb atoms and the applied spin states $\left|\downarrow\right\rangle =|42D_{\frac{5}{2}},m_{J}=\frac{5}{2}\rangle$, $\left|\uparrow\right\rangle =|44P_{\frac{3}{2}},m_{J}=\frac{3}{2}\rangle$ and $|F\rangle= |n_F F_{\frac{7}{2}},m_{J}=\frac{7}{2}\rangle$.
% we adjust the detuning, the strength of the laser, and the bond length to determine the coupling strength.
% \subsection{Periodic Configuration for Laser-assisted Interaction $J_{+-}/J_{--}$}\label{laddi configuration}
As mentioned in the main text, we adopt an 8-site periodic configuration for the on-site energy shift and Raman potential, enabling independent control of $J_{+-}/J_{--}$ on different bonds, as depicted in Fig.~\ref{periodic configuration}(a), the values of on-site energy shift $\Delta$ given in Table \ref{energy table}. In this configuration, LADDI along each bond is assisted by the Raman potential applied at the right end site, as illustrated in Fig.~\ref{periodic configuration}(b). For exchange $J_{+-}$ term, the bare dipole-dipole interaction strength between $\left|\uparrow\right\rangle_i \left|\downarrow \right\rangle_j$ and  $\left|\downarrow \right\rangle_i\left|\uparrow\right\rangle_j$ is $J_{d1}=C_3^{(1)}/r_{ij}^3$, where $C_3^{(1)}=528\,\mathrm{MHz\cdot \mu m^3}$ along $x,y$ bonds. We can choose $x,y$ bond length to be $a=4.7\,\mathrm{\mu m}$, so that $J_{d1}=5\,\mathrm{MHz}$. Assisting exchange process $J_{+-}$ with $|U_e/\Delta_e|=1/5$, a uniform $J_{+-}=-1\,\mathrm{MHz}$ can be achieved. Meanwhile, long-range $J_{+-}$ terms arise due to the long-range nature of dipole-dipole interaction. The subleading ones are bare $J_{d,+-}$ terms between sites with same energy shift $\Delta_i$. Such terms are suppressed in the 8-site periodic configuration because sites are well separated, and we find $J_{d,+-}/J_{+-}\leq 9.4\% $ that can be neglected.

\begin{table}[H]
\centering
\label{energy table}
\begin{tabular}{|p{1.7cm}<{\centering}|p{1.7cm}<{\centering}|p{1.7cm}<{\centering}|p{1.7cm}<{\centering}|p{1.7cm}<{\centering}|p{1.7cm}<{\centering}|p{1.7cm}<{\centering}|p{1.7cm}<{\centering}|}
\hline
 $\Delta_1$  &  $\Delta_2$ & $\Delta_3$ & $\Delta_4$  &  $\Delta_1^{\prime}$ & $\Delta_2^{\prime}$ & $\Delta_3^{\prime}$ & $\Delta_4^{\prime}$  \\ \hline
 $0\mathrm{MHz}$ &  $10\mathrm{MHz}$ & $40\,\mathrm{MHz}$ & $30\,\mathrm{MHz}$  & $45\,\mathrm{MHz}$ & $35\,\mathrm{MHz}$ & $5\,\mathrm{MHz}$ & $15\,\mathrm{MHz}$ \\ \hline
\end{tabular}
\caption{The on-site energy shift $\Delta$, as labelled in Fig.~\ref{periodic configuration}.}
\end{table}
For pairing $J_{--}$ term, the bare dipole-dipole interaction strength between $\left|\uparrow\right\rangle_i\left|F\right\rangle_j$ and $\left|\downarrow \right\rangle_i\left|\downarrow\right\rangle_j$ is $J_{d2}=C_3^{(2)}/r_{ij}^3$, where $C_3^{(2)}=556\,\mathrm{MHz\cdot \mu m^3}$ along $x,y$ bonds. The $x,y$ bond length is chosen as $a=4.7\,\mathrm{\mu m}$, we have $J_{d2}=5.4\,\mathrm{MHz}$. Raman potential with $|U_p/\Delta_p|=1/5.4$ and alternating sign gives $J_{--}=-1\,\mathrm{MHz}$ on $x$ bond and $J_{--}=+1\,\mathrm{MHz}$ on $y$ bond, respectively. The resonance condition of pairing process between sites $i$ and $j$ is controlled by the sum of their energy shifts $\Delta_i+\Delta_j$, so bond-selective laser-assistance is made possible. Finally, by combining uniform $J_{+-}$ and staggered $J_{--}$, we obtain alternating $J_x/J_y$ interaction in each row, with $J_x=J_y=-4\,\mathrm{MHz}$. This configuration ensures that the Raman potential $U_{p}$ exclusively induces the pairing term $J_{--}$ without giving rise to any additional effects, such as the exchange interaction $J_{+-}$.
Specifically, the exchange process $J_{+-}$ cannot be generated by $U_p$, because the transition $|\uparrow\rangle_i | \downarrow \rangle_j \rightarrow |F\rangle_i | \downarrow \rangle_j\rightarrow| \downarrow \rangle_i | \uparrow \rangle_j$ can not happen (the last process is off-resonant). Another possible side effect is long-range pairing term, since the resonance condition can be satisfied not only by two nearest neighbor sites, but also by two sites in two different spatial periods. These long-range pairing terms are also suppressed by 8-site periodic configuration, because they only occur between distant sites. They are bounded by $J^{\prime}_{--}/J_{--}\leq 5.2\%$ and can be neglected.

The $J_z$ term is generated from van der Waals interaction, and can be calculated from dipole-dipole interaction and energy level \cite{walker2008consequences}, described by $J_z=C_6/r^6$, where $r$ is the bond length. Given our choice of {\em quantization axis} in Fig. \ref{Fig1}(a), the $J_z$ along $z$ bond is has parameter $C_6=23.3\,\mathrm{GHz\cdot \mu m^6}$. By choosing the length of $z$ bond $b=4.2\,\mathrm{\mu m}$, we obtain $J_z=4\,\mathrm{MHz}$, so that $|J_z|=|J_{x,y}|$. Furthermore, $x,y$ bonds are aligned to {\em magic angle} of $J_z$ where $C_6=0$ as a result of cancellation between different vdW channels. After taking into account small variation of "magic angle" due to non-uniform energy shift as in Fig.~\ref{periodic configuration}(a), we can achieve $|J_z/J_{x,y}|<4\%$ on $x,y$ bonds.

% \subsection{Lifetime of Rydberg states}\label{lifetime}
The lifetime of Rydberg states are $\tau_{\uparrow}=124\,\mu\mathrm{s},\,\tau_{\downarrow}=63\,\mu\mathrm{s}$, and they are sufficiently large since $\tau\gtrsim \tau_0=60\,\mu\mathrm{s}=240J^{-1}$, where $J=|J_{x,y,z}|=4\,\mathrm{MHz}$. For completeness, we need to further consider the effect of Raman process on lifetime. For exchange process assisted by Raman potential $U_e$, the wavefunction on intermediate state $|6P_{\frac{3}{2}}\rangle$ would be $\Omega_{1,e}/\Delta_1$, so the lifetime will be prolonged by factor $(\Omega_{1,e}/\Delta_1)^{-2}$. Lifetime of $|6P_{\frac{3}{2}}\rangle$ alone is $\tau_{6P}=120\,\mathrm{ns}$, and to achieve lifetime longer than $\tau_0=60\,\mu\mathrm{s}$, we require $\Omega_{1,e}/\Delta_1<0.045$ that constrains Raman potential intensity $|U_e|=|\Omega_1 \Omega_e/\Delta_1|<0.045|\Omega_{1,e}|$. This is compatible with experimental realization, since $|U_e|\leq 6\,\mathrm{MHz}$ can be achieved at $|\Omega_{1,e}|\leq150\,\mathrm{MHz}$, which is appropriate for Rydberg states. Similarly, the pairing process couples to $|5D_{\frac{5}{2}}\rangle$ with lifetime $\tau_{5D}=238\,\mathrm{ns}$, and lifetime requirement imposes $\Omega_{2,p}/\Delta_2<0.063$. The target value $|U_p|\leq 17.4\,\mathrm{MHz}$ can be achieved at an appropriate laser intensity, $\Omega_{2,p}\leq 300\,\mathrm{MHz}$.

\section{Detection of the bulk spectra} \label{gap_detection_detail}

\begin{figure}[H]
	\begin{centering}
		\includegraphics[width=0.8\columnwidth]{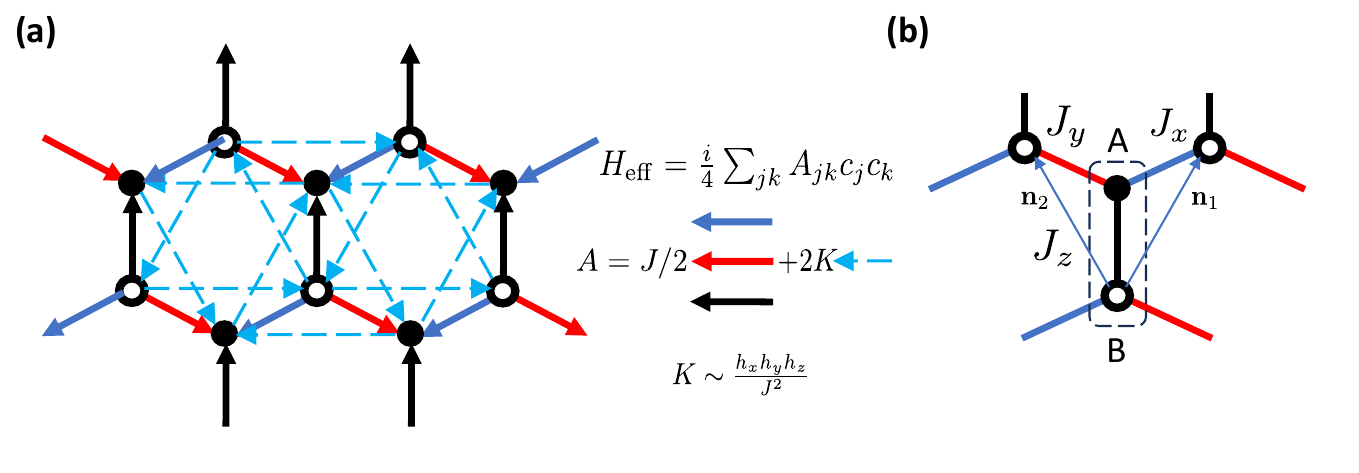}
		\par\end{centering}
\centering{}\caption{(a) Matrix element of $A$ in Eq. (\ref{majorana hamiltonian}). A plus sign is appended to the matrix element $A_{jk}$ when the arrow points from the site  $k$ to $j$, and a minus sign is used when the arrow points from the site $j$ to $k$. (b) Illustration of the Choice of  unit cell and basis vector.}\label{majorana hopping}
\end{figure}

Detection of Majorana bulk gap and chiral edge state rely on Majorana representation of the Kitaev spin liquid, by reducing spin-$1/2$ into free majorana fermions defined on lattice sites, and static $\mathbb{Z}_2$ gauge field on bonds \cite{kitaev2006anyons}. The key transformation is ${s_i^{\gamma}s_j^{\gamma}}_{\langle ij\rangle_{\gamma}}=-i u_{ij}c_i c_j/4$, where $c_i$ denotes Majorana operator on site $i$, and $u_{ij}$ denotes static gauge field along the bond connecting sites $i$ and $j$. According to Lieb's theorem \cite{lieb1994flux}, the ground state configuration of gauge field has zero flux in every plaquette, so we can fix $u_{ij}=1$ for $i\in A$ sublattice and $j\in B$ sublattice. Changing gauge field configuration is prohibited by a flux gap $\sim J$, so we can safely study low-energy physics in ground state flux-free sector in the following. Moreover, the term ${s_i^{\gamma}s_j^{\gamma}}_{\langle ij\rangle_{\gamma}}=-i c_i c_j/4\,(i\in A,\,j\in B)$ is the minimal perturbation that excites Majorana fermion without changing gauge field configuration, so it serves as probe of Majorana fermions. With the magnetic field, the effective Hamiltonian of Kitaev spin liquid in flux-free sector is given by \cite{kitaev2006anyons}
\begin{equation}\label{majorana hamiltonian}
H_{\mathrm{eff}}=\frac{i}{4}\sum_{jk}A_{jk}c_{j}c_{k},
\end{equation}
where $c_j$ denotes Majorana operator on site $j$, and matrix elements $A_{jk}$ are set as illustrated in Fig. \ref{majorana hopping}(a), with a plus (minus) sign in front if the arrow goes from $k$ to $j$ ($j$ to $k$). The next-nearest-neighbor exchange term $K$ arises from the third-order perturbation of the magnetic field,  given by $K\sim h_x h_y h_z /J^2$, and this term vanishes in the absence of a magnetic field.

With the choice of unit cell and basis vector as shown in Fig. \ref{majorana hopping}(b), we define a complex fermion in each unit cell $s$, $f_s=(c_{s,A}+ic_{s,B})/2$. The Hamiltonian of $f$ fermion contains exchange and pairing terms,
\begin{equation}
    H=\frac{1}{2}\sum_{\mathbf{q}}2\mathrm{Re}s_{\mathbf{q}}(f_{\mathbf{q}}^{\dagger}f_{\mathbf{q}}-f_{\mathbf{-q}}f_{\mathbf{-q}}^{\dagger})+2i\mathrm{Im}s_{\mathbf{q}}(f_{\mathbf{-q}}f_{\mathbf{q}}-f_{\mathbf{q}}^{\dagger}f_{\mathbf{-q}}^{\dagger})+\Delta_{\mathbf{q}}(f_{\mathbf{q}}^{\dagger}f_{\mathbf{-q}}^{\dagger}+f_{\mathbf{-q}}f_{\mathbf{q}}),
\end{equation}
where $s_{\mathbf{q}}=(J_z+J_x e^{i\mathbf{q}\cdot\mathbf{n}_1}+J_y e^{i\mathbf{q}\cdot\mathbf{n}_2})/4$ and $\Delta_{\mathbf{q}}=4K(\sin(\mathbf{q\cdot n}_{1})-\sin(\mathbf{q\cdot n}_{2})+\sin(\mathbf{q\cdot n}_{2}-\mathbf{q\cdot n}_{1}))$. By performaing the Bogoliubov transformation $f_{\mathbf{q}}=\cos\theta_{\mathbf{q}}d_{\mathbf{q}}+i\sin\theta_{\mathbf{q}}e^{i\phi_{\mathbf{q}}}d_{-\mathbf{q}}^{\dagger}$, where $\tan\phi_{\mathbf{q}}=\Delta_{\mathbf{q}}/(2\mathrm{Im}s_{\mathbf{q}})$, $\tan2\theta_{\mathbf{q}}=\sqrt{\Delta_{\mathbf{q}}^{2}+4\mathrm{Im}s_{\mathbf{q}}^{2}}/(2\mathrm{Re}s_{\mathbf{q}})$, the Hamiltonian is diagonalized as
\begin{equation}
    H=\sum_{\mathbf{q}}\epsilon_{\mathbf{q}}(d_{\mathbf{q}}^{\dagger}d_{\mathbf{q}}-\frac{1}{2}),\,\epsilon_{\mathbf{q}}=\sqrt{4|s_{\mathbf{q}}|^{2}+\Delta_{\mathbf{q}}^{2}}.
\end{equation}
The ground state is the vacuum of $d$ fermion. The perturbation $\Delta H=\sum_{\langle ij\rangle_{\gamma=x,y}} \Delta J  s^{\gamma}_{i} s^{\gamma}_{j}$ to detect gap can be rewritten as
\begin{equation}
\begin{aligned}
\Delta H= & \sum_{\mathbf{q}}(\mathrm{Re}h_{\mathbf{q}}\cos2\theta_{\mathbf{q}}+\mathrm{Im}h_{\mathbf{q}}\sin2\theta_{\mathbf{q}}\cos\phi_{\mathbf{q}})d_{\mathbf{q}}^{\dagger}d_{\mathbf{q}}  \\
&+\left[i\mathrm{Im}h_{\mathbf{q}}(\cos^{2}\theta_{\mathbf{q}}-\sin^{2}\theta_{\mathbf{q}}e^{-2i\phi_{\mathbf{q}}})-i\mathrm{Re}h_{\mathbf{q}}\sin2\theta_{\mathbf{q}}e^{-i\phi_{\mathbf{q}}}\right]d_{\mathbf{-q}}d_{\mathbf{q}}+\mathrm{H.c.}, \\
h_{\mathbf{q}}=&\Delta J(e^{i\mathbf{q\cdot n}_{1}}+e^{i\mathbf{q\cdot n}_{2}})/4.
\end{aligned}
\end{equation}
Then $I(\omega)$ defined in main text is given by
\begin{equation}\label{magnetic gap response}
    I(\omega)=-4\pi\sum_{\mathbf{q}}\delta(\omega-2\epsilon_{\mathbf{q}})\left|i\mathrm{Im}h_{\mathbf{q}}(\cos^{2}\theta_{\mathbf{q}}-\sin^{2}\theta_{\mathbf{q}}e^{-2i\phi_{\mathbf{q}}})-i\mathrm{Re}h_{\mathbf{q}}\sin2\theta_{\mathbf{q}}e^{-i\phi_{\mathbf{q}}}\right|^{2},\,\omega>0,
\end{equation}
which is proportional to density of states $g(\omega/2)$, and $I(-\omega)=-I(\omega)$. This expression (\ref{magnetic gap response}) reduce to Eq. (\ref{gap response}) if magnetic field is zero, i.e., the next-nearest hopping $K=0$. The key feature of of (\ref{magnetic gap response}) is that $I(\omega)\propto g(\omega/2)$, where $g(\omega)=\sum_{\mathbf{q}}\delta(\omega-\epsilon_{\mathbf{q}})$ is the single-particle density of states. This can be understood intuitively, since perturbation $\Delta H=\sum_{\langle ij\rangle_{\gamma=x,y}} \Delta J  s^{\gamma}_{i} s^{\gamma}_{j}$ couples to two Majorana fermions $c_i c_j$, so the excitation energy is doubled compared to single-particle case.

\section{Detection of the edge modes} \label{edge_detection_detail}

\begin{figure}[H]
	\begin{centering}
		\includegraphics[width=0.5\columnwidth]{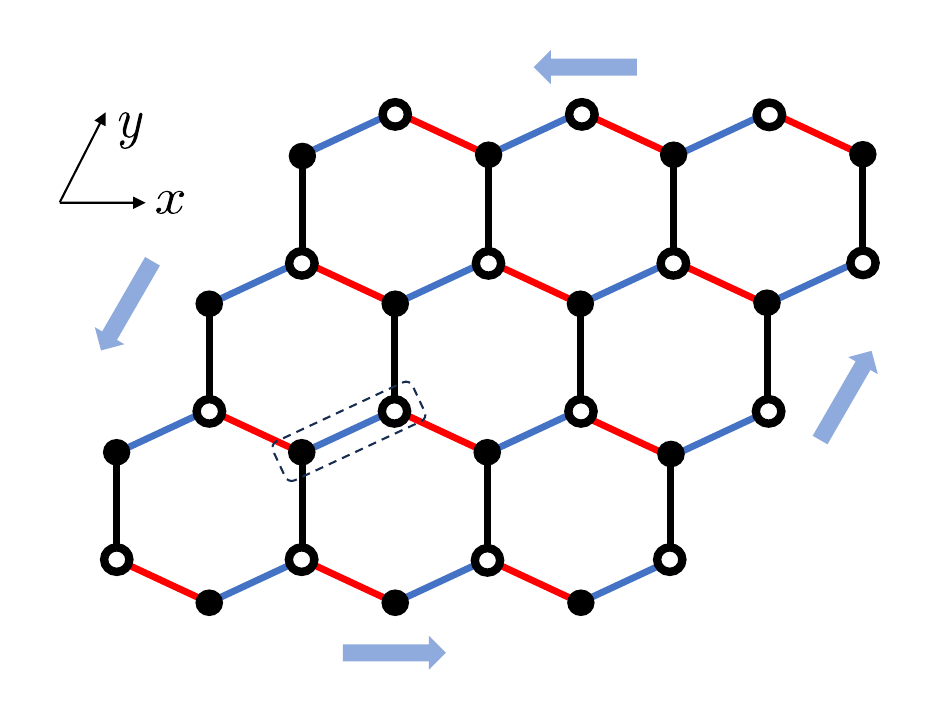}
		\par\end{centering}
	\centering{}\caption{Open boundary geometry with blue arrows indicating the counter-clockwise propagating edge mode. Dashed box indicates the unit cell definition in Fig. \ref{supfig-evolution}.}\label{obc}
\end{figure}

In this section, we present more details of edge state detection of Kitaev chiral spin liquid. First, we explain the topological origin of the edge state. Chern number can be defined on non-interacting Majorana band \cite{kitaev2006anyons}, based on Majorana representation shown in Sec. \ref{gap_detection_detail}, taking values $\pm 1$ in the topological case. However, the topology of Kitaev chiral spin liquid can be understood in a more heuristic way. Consider two copies of the model, with Majorana fermion on $i$ site denoted by $c_i$ and $c_i^{\prime}$. Then we combine these two copies two obtain a system of free complex fermion $\alpha_i$, where $\alpha_i=(c_i+ic_i^{\prime})/2$ for $i\in A$ sublattice, and $\alpha_i=i(c_i+ic_i^{\prime})/2
$ for $i\in B$ sublattice. Hamiltonian of $\alpha$ fermion is the Haldane model of QAHE \cite{haldane1988model}, with next-nearest hopping phase $\phi=\pi/2$. It is well known that Haldane model of QAHE has Chern number $C=\pm 1$, and consequently has one chiral edge mode of complex fermion. Since two copies are independent, each of them must possess one chiral edge mode of Majorana fermion. In the calculation related to edge state, we choose the open boundary geometry shown in Fig. \ref{obc}. The Chern number is chosen to be $C=1$, so the bottom (top) edge mode is right (left) moving as indicated by blue arrow in Fig. \ref{obc}.
% We can
% label a site i by i = (l, m) with two positive integers, l and m, which satisfy $1 \le l \le 2L_x$ and $1 \le m \le L_y$ with the length $2L_x$ of the cylinder and the length $L_y$ of the
% circumference of the cylinder.

\begin{figure}[htbp]
\begin{centering}
\includegraphics[width=\columnwidth]{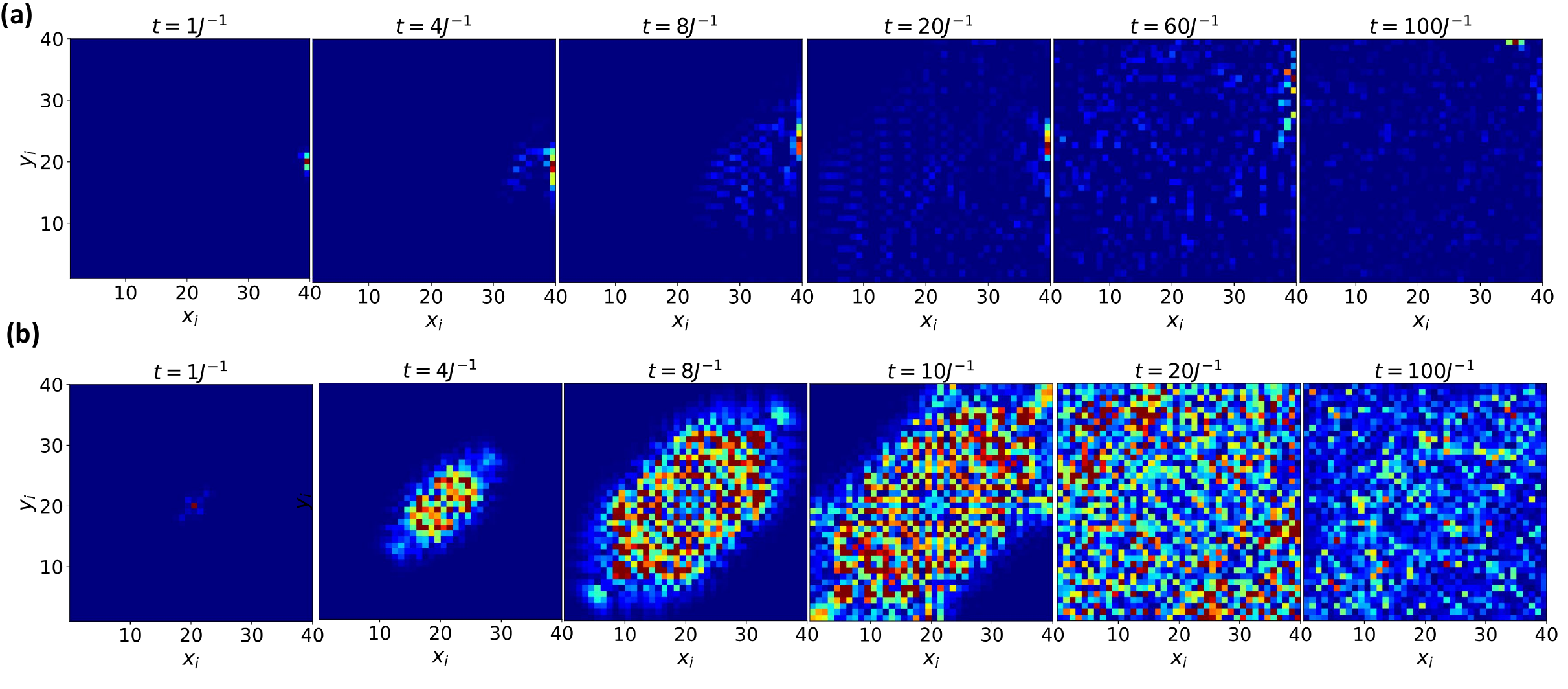}
\par\end{centering}
\centering{}\caption{Time evolution of the excited Majorana fermions in both edge and bulk. (a) A short-time pulse is applied at the system's edge, resulting in the excitation of  Majorana fermions which subsequently undergo dynamic evolution. The majority of these Majoran fermions are excited to the chiral edge states, displaying unidirectional motion.
(b) A pulse applied at the system's center leads to the excitation of Majorana fermions primarily in the bulk states. The excitations  subsequently spread throughout the entire system.}\label{supfig-evolution}
\end{figure}

Our first strategy images the unidirectional particle flow at edge carried by chiral edge mode. We employ a short-time pulse at one specific bond to induce Majorana wavepacket and observe their subsequent propagation. The pulse strength is chosen as  $\delta J_x=0.5$, and its duration is taken as $\delta t =0.1 J_x^{-1}$. This pulse couples to ${s_i^x s_j^x}_{\langle ij\rangle_x}=-i c_i c_j/4$, enabling the excitation of Majorana fermions to higher energy states.
To visualize the propagation of the excited Majorana fermions in both the edge and bulk regions, we present Fig.~\ref{supfig-evolution}.  Here, each site within the system is denoted by the label $(x_i,y_i)$, with one lattice site representing a unit cell, as shown by dashed box in Fig. ~\ref{obc}. In Fig.~\ref{supfig-evolution}(a), we initially apply the pulse at the edge, precisely at position $(x_i=40, y_i=20)$, until time $t=0.1J^{-1}$. As depicted in the first figure, this pulse excites a portion of particles. Because the pulse can be decomposed to the superposition of many different frequencies, some Majorana fermions are actually excited to the bulk states and diffuse within the bulk region, as observed at time $t=0.4J^{-1}$ and $t=0.8J^{-1}$. However, the majority of excitations are excited towards the chiral edge states and start moving along the edge in the upward direction. In Fig.~\ref{supfig-evolution}(b), we apply the pulse at the center of the system ($x_i=y_i=20$). The Majorana fermions get excited to the bulk states and then spread throughout the entire system. Notably, the diffusion along one diagonal direction occurs much more rapidly than in other directions.  This behavior arises due to our lattice label naturally  exhibits anisotropic properties, which converts the honeycomb lattice into a square lattice.

\begin{figure}[H]
\begin{centering}
\includegraphics[width=\columnwidth]{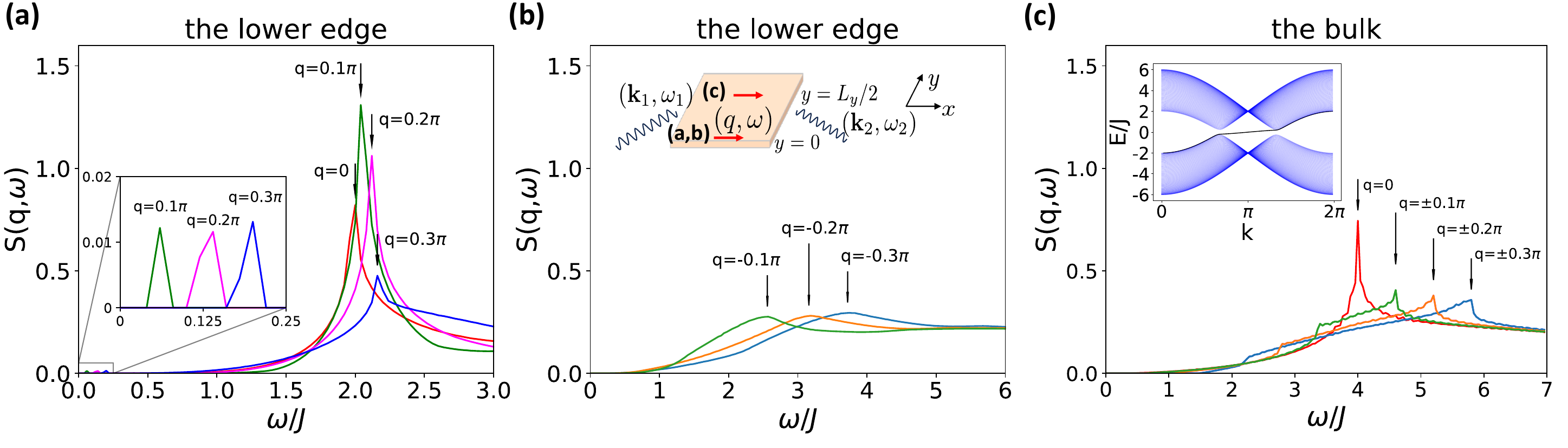}
	\par\end{centering}
\centering{}\caption{Structure factor measured on lower edge and in the bulk. (a) Structure factor measured on lower edge with positive momentum change $q$, exhibiting resonant peaks with positive group velocity $v_f=\Delta \omega_{\mathrm{peak}}/\Delta q>0$ due to chiral edge mode. (b) Structure factor measured on lower edge, but with negative momentum change $q$, with no resonant peak is observed. (c) Structure factor measured in the bulk with symmetric profile for $\pm q$. Inset: Schematic of light Bragg scattering at $y=0$ edge, and the energy spectrum with lower edge state in black line.}\label{sup edge bragg}
\end{figure}

In the detection of the chiral edge state with light Bragg scattering, we measure the absorption of
the lasers to  obtain the dynamical structure factor $S(q, \omega)$, which is defined as
\begin{equation}
\begin{aligned}
S(q, \omega) &=\sum_{k_{x}} \left[1-f(E^{(f)}_{k_{x}+q}) \right] f(E^{(i)}_{k_{x}})  \\
&\times \left| \langle \Psi^{(f)}_{k_{x}+q} | \Delta H^{\prime} |  \Psi^{(i)}_{k_{x}} \rangle \right|^2 \delta \left[ \hbar \omega - E^{(f)}_{k_{x}+q} + E^{(i)}_{k_{x}} \right].
\end{aligned}
\end{equation}
with $|\Psi^{(i)}_{k_{x}} \rangle$ ($|\Psi^{(f)}_{k_{x}+q} \rangle$) the initial (final) state before (after) scattering, and $f(E)$ the fermi distribution function where negative energy state occupied in energy spectrum shown in Fig. \ref{Fig4}(d).
Then we provide supplementary information in Fig. \ref{sup edge bragg}. When the Raman perturbation is applied at $y=0$ edge and the momentum change $q$ is positive, a series of resonant peaks with $v_f=\Delta \omega_{\mathrm{peak}}/\Delta q>0$ are observed in Fig. \ref{sup edge bragg}(a), arising from the chiral dispersion of edge mode. In sharp contrast, edge structure factor for negative momentum change $q$ (Fig. \ref{sup edge bragg}(b)) does not exhibit any  resonant peak, because transition to occupied negative energy chiral edge mode is prohibited. If the structure factor is measured in the bulk (Fig. \ref{sup edge bragg}(c)), the result is symmetric for $\pm q$, because they only involve bulk states that distribute symmetrically in momentum. Such contrast between asymmetry on the edge and symmetry in the bulk further indicates existence of chiral mode localized at edge.

\end{document}